\documentclass[prd,preprint,showpacs,eqsecnum,floatfix]{revtex4}

\usepackage{amsfonts}
\usepackage{amssymb}
\usepackage{graphicx}
\usepackage{pstricks}
\usepackage{axodraw}

\newcommand{\be}{\begin{equation}}
\newcommand{\bea}{\begin{eqnarray}}
\newcommand{\ee}{\end{equation}}
\newcommand{\eea}{\end{eqnarray}}
\newcommand{\bpi}{\begin{picture}}
\newcommand{\bce}{\begin{center}}
\newcommand{\epi}{\end{picture}}
\newcommand{\ece}{\end{center}}

\newcommand{\de}{d^{-1}}

\newcommand{\pci}{p_i}

\newcommand{\ksm}{k\hspace{-0.24cm}/}

\newcommand{\psm}{p\hspace{-0.18cm}/}
\newcommand{\psmp}{p\hspace{-0.14cm}/}
\newcommand{\D}{\displaystyle}

\def\g{{\rm I}\hspace{-0.07cm}\Gamma}

\begin{document}

\title{Pinch technique self-energies and vertices\\[0.2cm] 
to all  orders in perturbation theory}

%\date{\today}

\author{Daniele Binosi}
\author{Joannis Papavassiliou}
\affiliation{Departamento de F\'\i sica Te\'orica and IFIC, Centro Mixto, 
Universidad de Valencia-CSIC,
E-46100, Burjassot, Valencia, Spain}

\email{Daniele.Binosi@uv.es; Joannis.Papavassiliou@uv.es}

\begin{abstract}

The all-order  construction of  the pinch technique  gluon self-energy
and  quark-gluon vertex  is presented  in detail  within the  class of
linear covariant gauges.  The main ingredients in our analysis are the
identification of a special Green's function, which serves as a common
kernel to all  self-energy and vertex diagrams, and  the judicious use
of  the Slavnov-Taylor identity  it satisfies.   In particular,  it is
shown  that the  ghost-Green's  functions appearing  in this  identity
capture  precisely the  result  of the  pinching  action at  arbitrary
order.  By  virtue  of   this  observation  the  construction  of  the
quark-gluon vertex  becomes particularly  compact.  It turns  out that
the aforementioned ghost-Green's functions  play a crucial role, their
net effect being the non-trivial modification of the ghost diagrams of
the quark-gluon vertex  in such a way as  to reproduce dynamically the
characteristic ghost sector of the background field method.  The gluon
self-energy  is also constructed  following two  different procedures.
First, an  indirect derivation  is given, by  resorting to  the strong
induction  method  and  the   assumption  of  the  uniqueness  of  the
$S$-matrix.  Second,  an explicit construction based  on the intrinsic
pinch  technique  is   provided,  using  the  Slavnov-Taylor  identity
satisfied  by  the  all-order  three-gluon vertex  nested  inside  the
self-energy   diagrams. 
The   process-independence   of  the   gluon
self-energy  is also  demonstrated,  
by  using gluons  instead  of quark  as
external   test   particles,   and  identifying   the
corresponding  kernel  function,   together  with  its  Slavnov-Taylor
identity.  Finally,  the  general  methodology for  carrying  out  the
renormalization  of the  resulting  Green's functions  is outlined, 
and various open questions are briefly discussed.

\end{abstract}


\preprint{FTUV-03-0114}
\preprint{IFIC-02-20}

\maketitle

\section{Introduction}

When quantizing  gauge theories in  the continuum 
one must invariably resort to an appropriate gauge-fixing procedure in
order   to  remove  redundant   (non-dynamical)  degrees   of  freedom
originating    from    the   gauge    invariance    of   the    theory
\cite{Wilson:1974sk}.    
Thus,  one  adds   to  the   gauge  invariant
(classical) Lagrangian  ${\cal L}_{\rm I}$ a  gauge-fixing term ${\cal
L}_{\rm GF}$, which allows for the consistent derivation of
Feynman  rules.  At  this point  a new  type of  redundancy  makes its
appearance, this time at the level of the building blocks defining the
perturbative  expansion. In  particular, individual  off-shell Green's
functions  ($n$-point  functions) carry  a  great  deal of  unphysical
information,  which disappears when  physical observables  are formed.
$S$-matrix elements, for example,  are independent of the gauge-fixing
scheme  and  parameters  chosen  to  quantize  the  theory,  they  are
gauge-invariant  (in  the sense  of  current  conservation), they  are
unitary  (in  the sense  of  conservation  of  probability), and  well
behaved at high energies.  On  the other hand Green's functions depend
explicitly (and generally non-trivially) on the gauge-fixing parameter
entering  in the  definition of  ${\cal L}_{\rm  GF}$, they  grow much
faster than  physical amplitudes at  high energies (e.g.  they grossly
violate   the   Froissart-Martin   bound \cite{Froissart:ux}),
  and   display   unphysical
thresholds.   Last but  not  least,  in the  context  of the  standard
path-integral  quantization  by  means  of the  Faddeev-Popov  Ansatz,
Green's functions satisfy complicated Slavnov-Taylor identities (STIs)
\cite{Slavnov:1972fg}  involving ghost  fields, instead  of  the usual
Ward  identities (WIs)  generally associated  with the  original gauge
invariance.

The  above observations imply  that in  going from  unphysical Green's
functions to physical  amplitudes, subtle field theoretical mechanisms
are at work, implementing vast cancellations among the various Green's
functions. Interestingly enough,  these cancellations may be exploited
in   a   very   particular   way   by   the   Pinch   Technique   (PT)
\cite{Cornwall:1982zr,Cornwall:1989gv,Papavassiliou:1990zd,
Degrassi:1992ue}: 
a  given physical amplitude is
reorganized  into  sub-amplitudes,   which  have  the  same  kinematic
properties       as       conventional       $n$-point       functions
(self-energies,vertices,boxes) but, in addition, they are endowed with
important   physical   properties.    This   has   been   accomplished
diagrammatically, at the one- and two-loop level, by recognizing that
{\it longitudinal momenta} circulating  inside vertex and box diagrams
generate (by ``pinching''  out internal fermion lines) propagator-like
terms.  The  latter are reassigned to  conventional self-energy graphs
in order to give rise  to effective Green's functions which manifestly
reflect the properties generally associated with physical observables.
In  particular,  the  PT  Green's  function  are  independent  of  the
gauge-fixing  scheme  and parameters  chosen  to  quantize the  theory
($\xi$ in covariant  gauges, $n_\mu$ in axial gauges,  etc.), they are
gauge-invariant, {\it  i.e.}, they satisfy  simple tree-level-like WIs
associated with the gauge  symmetry of the classical Lagrangian ${\cal
L}_{\rm  I}$,  they  display  only  {\it  physical  thresholds},  and,
finally, they are well behaved at high energies.

There are two basic questions that are of particular relevance in this 
context: ({\it i}) what are the conceptual and phenomenological  
advantages of being able to work with such special Green's 
functions, and ({\it ii}) how to achieve their systematic construction 
to all orders  in perturbation  theory.
Before turning to the second question, which constitutes the 
main thrust of this paper, we will briefly discuss the 
first one, in an attempt to physically motivate the  
technical presentation that will follow \cite{dis}. 

\begin{itemize}
 
\item 
{\it   QCD  effective    charge:}   
The unambiguous extension  of the  concept of  the 
gauge-independent, renormalization
group invariant,  and process independent 
\cite{Grunberg:1992mp} effective charge  
from QED to
QCD \cite{Cornwall:1976ii,Cornwall:1982zr},
is of special  interest for several reasons 
\cite{Watson:1996fg}.  
The PT  construction of this quantity 
 accomplishes
the explicit identification  of the conformally-(in)variant subsets of
QCD  graphs  
\cite{Brodsky:1982gc},   
usually assumed in the field of renormalon calculus 
\cite{Mueller:1992xz}.  
In
addition, the PT effective charge  can serve as the natural scheme for
defining the  coupling in the proposed  ``event amplitude generators''
based on  the the light-cone formulation  of QCD \cite{Brodsky:2001ha}.

\item
{\it  Breit-Wigner resummations,  
resonant  transition amplitudes, unstable particles:}
The  Breit-Wigner  
procedure used for regulating  the physical singularity 
appearing in the  vicinity  of resonances ($\sqrt{s}\sim M$) 
is equivalent
to a {\it reorganization}  of the perturbative series 
\cite{Veltman:th}.  
In particular,  the Dyson 
summation of the self-energy $\Pi(s)$ amounts to 
removing a particular piece
from  each order  of the  perturbative expansion,  since from  all the
Feynman graphs contributing to a given order $n$ one only picks 
the part
which contains $n$  self-energy bubbles  $\Pi(s)$, 
and  then one takes  $n \to
\infty$.  
Given  that non-trivial cancellations  involving the various
Green's function is  generally taking place at any  given order of the
conventional perturbative  expansion, the act of removing  one of them
from each order  may distort those cancellations; 
this  is indeed what
happens  when constructing  non-Abelian {\it running  widths}.   
The  PT  ensures   that  all  unphysical
contributions contained inside $\Pi(s)$ have been identified and properly
discarded, {\it before} $\Pi(s)$ undergoes resummation 
\cite{Papavassiliou:1995fq}.

\item 
{\it Off-shell form-factors:}
In non-Abelian theories their proper definition poses in general 
problems related to the gauge invariance \cite{Fujikawa:fe}. 
Some representative cases have been the 
magnetic dipole and electric 
quadrupole moments of the $W$ \cite{Papavassiliou:ex}, 
the top-quark magnetic moment \cite{Papavassiliou:1993qe}, 
and the neutrino charge radius \cite{Bernabeu:2000hf}. 
The PT allows for an unambiguous   
definition of such quantities; most notably, 
the gauge-independent, renormalization-group- invariant, 
and target-independent neutrino charge radius 
constitutes a genuine {\it physical} observable,   
since it can be 
extracted (at least in principle) from experiments 
\cite{Bernabeu:2002nw}.

\item
{\it  Schwinger-Dyson  equations:}  This infinite
system of  coupled non-linear integral 
equations  for all Green's  functions of
the  theory is inherently  non-perturbative  and can
accommodate phenomena  such as  chiral symmetry breaking  and dynamical
mass  generation.   In  practice one  is  severely
limited  in their  use,  and a  self-consistent  truncation scheme  is
needed. The main  problem in this context is  that the Schwinger-Dyson
equations are  built out  of gauge-dependent Green's  functions; since
the  cancellation  mechanism  is  very subtle,  involving  a  delicate
conspiracy of terms  from {\it all orders}, a  casual truncation often
gives   rise   to   gauge-dependent  approximations   for   ostensibly
gauge-independent quantities \cite{Cornwall:1974vz,Marciano:su}. 
The role of the PT in this problem is to
(eventually)  trade   the  conventional  Schwinger-Dyson   series  for
another,  written  in terms  of  the  new, gauge-independent  building
blocks \cite{Cornwall:1982zr,Mavromatos:1999jf,Sauli:2002tk}. 
The upshot  of this program would then
be  to truncate this  new series,  by keeping  only a  few terms  in a
``dressed-loop'' expansion, and maintain exact gauge-invariance, while
at the same time accommodating non-perturbative effects.

\item 
Other interesting applications include 
the gauge-invariant formulation of the $\rho$ parameter 
at one-\cite{Degrassi:1993kn} 
and two-loops \cite{Papavassiliou:1995hj}, 
various finite temperature calculations 
\cite{Nadkarni:1988ti}, 
a novel approach to the comparison of electroweak data with 
theory \cite{Hagiwara:1994pw},
resonant CP violation \cite{Pilaftsis:1997dr},
the construction of the two-loop PT quark self-energy \cite{Binosi:2001hy}, 
and more recently the issue of particle mixings
\cite{Yamada:2001px}.    

\end{itemize}

After this digression we return to the main 
issue to be addressed in this paper, namely the 
generalization of the PT to all orders. 
The original one-loop \cite{Cornwall:1982zr} and 
two-loop \cite{Papavassiliou:2000az}
PT calculations consist in carrying out algebraic manipulations 
  inside individual box- and vertex-diagrams, 
following well-defined rules. In particular 
one tracks down the rearrangements 
induced when the action of 
(virtual) longitudinal momenta ($k$) on the bare vertices of 
diagrams trigger elementary WIs. 
Eventually a WI of the form 
$ k_{\mu}\gamma^{\mu} = S^{-1}(\ksm + \psm)- \,S^{-1}(\psm)$
 will give rise to 
propagator-like parts, by removing (pinching out) the 
internal bare fermion propagator $S(\ksm + \psm)$ \cite{WI}.  
Depending on the order and topology 
of the diagram under consideration, the  
final  WI may be activated 
immediately, as happens at one-loop \cite{Cornwall:1982zr,Cornwall:1989gv}, 
or as the final outcome of a sequential 
triggering of intermediate WIs, as happens at two-loops  
\cite{Papavassiliou:2000az}. 
The propagator-like contributions so obtained 
are next reassigned to the usual gluon self-energies,  
giving rise to the PT  gluon self-energy.  
The  longitudinal momenta  responsible  for these
rearrangements stem either from the bare gluon propagators or from the
pinching  part appearing
in  the  characteristic  decomposition  of the  tree-level
(bare) three-gluon   vertex.

\begin{figure}[!t]
\includegraphics[width=8cm]{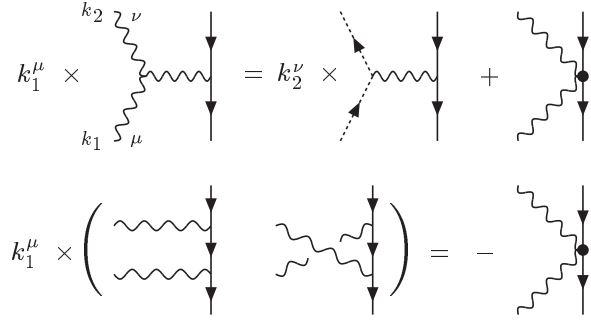}
\caption{\label{fig:1} The tree-level version 
of the fundamental $s$-$t$ channel cancellation.}
\end{figure}

As  we will  explain in  detail  in what  follows, the  aforementioned
rearrangements  are  but lower-order  manifestations  of a  fundamental
cancellation taking place between  graphs of distinct kinematic nature
when computing the divergence  of the four-point function $A_{\mu}^a\,
A_{\nu}^b\, q^i\, \bar{q}^j$, with the gluons $A^a_{\mu}$, $A^b_{\nu}$
off-shell,   and  the  quarks   $q^i$,  $\bar{q}^j$   on-shell).   The
importance of  this particular amplitude has first  been recognized in
the third paper of \cite{Papavassiliou:1995fq},  
where   the  tree-level  version   of  this
cancellation  was  considered: when  the  $s$-channel and  $t$-channel
diagrams  of Fig.1  ({\it i.e.},  the tree-level  contribution  to the
amplitude $A_{\mu}^a\, A_{\nu}^b\, q^i\, \bar{q}^j$) are contracted by
a  common  longitudinal momentum,  one  obtains  from  either graph  a
common,  propagator-like part,  which eventually  cancels  against the
other.   These parts  display  the characteristic  feature that,  when
depicted  by  means  of  Feynman  diagrams,  they  contain  unphysical
vertices (Fig.\ref{fig:1}), {\it i.e.}, vertices which do not exist in
the original Lagrangian \cite{com1} ; they correspond precisely to the
``pinch parts'' mentioned above.  It turns out that the aforementioned
four-point function constitutes a common kernel to all self-energy and
vertex  diagrams  appearing in  the  process  $q^m  \bar{q}^n \to  q^i
\bar{q}^j$.  As  has  been  shown  in  a  recent  brief  communication
\cite{Binosi:2002ft} the  judicious exploitation of the  STI that this
Green's function satisfies allows  for the all-order generalization of
the ($S$-matrix) PT procedure.   We emphasize that the method outlined
in \cite{Binosi:2002ft},  which will be  explained in great  detail in
the present paper, does not constitute  a new definition of the PT, but
rather  a  new,   far  more  expeditious  way  of   carrying  it  out.
Essentially  one is trading  off  the  tree-level WIs  employed  in  the
algebraic manipulations  of individual  Feynman graphs --  following a
well-defined, albeit cumbersome procedure  which clearly does not lend
itself  for  an  all-order  construction  --  for  the  formally  more
complicated,  but  operationally far  more  efficient, all-order  STIs
imposed  on  the  (kernel)   four-point  function  by  the  underlying
Becchi-Rouet-Stora-Tyutin (BRST) symmetry \cite{Becchi:1976nq}.

In this paper  we will focus on the following three main points:
First, we will present in detail the various technical
aspects of the all-order construction presented in 
\cite{Binosi:2002ft}, and further elaborate on the 
crucial role of the STI satisfied by  
the relevant four-point function.
Second, we present the all orders generalization  of the 
{\it intrinsic} PT procedure, which will allow for the explicit
construction of the all-order PT gluon self-energy.       
Finally, we will show that the construction of the 
PT two-point function is 
{\it universal}  (process-independence); this will be accomplished 
by studying the STI of the Green's function 
$A^a_{\mu} \, A^b_{\nu}\, A^{e_1}_{\sigma_1}
\, A^{e_2}_{\sigma_2}$, which appears in the  
alternative on shell processes 
$q^m\, \bar{q}^n \to A^{e_1}_{\sigma_1} \,A^{e_2}_{\sigma_2}$ and 
$A^{d_1}_{\rho_1}\, A_{\rho_2}^{d_2}\to  A^{e_1}_{\sigma_1} 
\,A^{e_2}_{\sigma_2}$.

The  rest of  the  paper is  organized  as follows.  In Section~II  we
outline  the  general framework  of  the  $S$-matrix  PT, isolate  the
aforementioned particular Green's  function which constitutes a kernel
to  all  higher  order diagrams,  and  derive  in  detail the  STI  it
satisfies.  In  Section~III  we   explain  in  detail  why  the  usual
fundamental  PT cancellations  are in  fact  encoded in  this STI,  an
observation  which  eventually   makes  the  all-order  generalization
possible.  In  Section~IV  we   carry  out  explicitly  the  all-order
construction  of the PT  gluon--quarks--anti-quark
vertex. Section~V  is  dedicated to  the  explicit all-order construction
of the PT gluon self-energy, following the ``intrinsic PT'' algorithm.
In  Section~VI we address
the issue  of the universality of the  PT gluon self-energy,
proving in a direct way that it is process-independent.
In Section~VII  we discuss the general methodology 
that must be followed in order to carry out the renormalization 
of the effective PT Green's functions.  
Finally,  in  Section~VIII  we present  our conclusions.

\section{The four-point kernel and its Slavnov-Taylor Identity}

In this section  we will explain how the  four-point function $A^a_{\mu}
A^b_{\nu}  \,  q^i  \,\bar{q}^j$  acquires  its central  role  
in  the  PT construction, 
and  will derive  in detail the  STI that  it satisfies.
This  STI  will  be  instrumental  in the  study  of  the  fundamental
cancellations taking place between the (all-order) two- and three-point
functions   embedded  into   $S$-matrix  elements,   leading   to  the
generalization of the $S$-matrix PT to all-orders.

Let us focus on the  $S$-matrix
element  for the  quark--anti-quark elastic  scattering  process $
q^m(r_1) \bar q^n(r_2)\to q^i(p_1) \bar q^j(p_2)$ in  QCD, typically 
considered in the PT construction. 
We set $q=r_2-r_1=p_2-p_1$,
with $s=q^2$ the  square of the momentum transfer.
The longitudinal momenta responsible  for triggering the
kinematical rearrangements characteristic of the PT 
stem either
from   the  bare  gluon   propagator, $\Delta_{\mu\nu}^{[0]}(k)$,
which in the covariant renormalizable gauges assumes the form
\be
\Delta_{\mu\nu}^{[0]}(k) = 
-\frac{\D i}{\D k^2}
\left[\ g_{\mu\nu} - (1-\xi) \frac{\D k_\mu
k_\nu}{\D k^2}\right] \,,
\label{GluProp}
\ee 
or   from  the  {\it external}
tree-level three-gluon vertices, {\it i.e.}, the vertices where the
physical momentum transfer $q$ is entering \cite{EXVER}.
The latter, to be denoted by
$\Gamma^{eab\,[0]}_{\alpha\mu\nu}(q,k_1,k_2)$, is given by the
following manifestly Bose-symmetric expression
(all momenta are incoming, {\it i.e.}, $q+k_1+k_2 = 0$)
\bea
& &\Gamma^{eab\,[0]}_{\alpha\nu\mu}(q,k_1,k_2)=g f^{eab}
\Gamma_{\alpha \mu \nu}^{[0]}(q,k_1,k_2), \nonumber \\
& & \Gamma_{\alpha \mu \nu}^{[0]}(q,k_1,k_2)= 
(q-k_1)_{\nu}g_{\alpha\mu} + (k_1-k_2)_{\alpha}g_{\mu\nu}
 + (k_2-q)_{\mu}g_{\alpha\nu}.
\label{tgv}
\eea
$\Gamma_{\alpha \mu \nu}^{[0]}(q,k_1,k_2)$
may be then split into two parts \cite{Cornwall:1976ii}
\be
\Gamma_{\alpha \mu \nu}^{[0]}(q,k_1,k_2) 
=\Gamma_{\alpha \mu \nu}^{{\rm F}}(q,k_1,k_2)+
\Gamma_{\alpha \mu \nu}^{{\rm P}}(q,k_1,k_2),
\label{decomp}
\ee
with 
\bea
\Gamma_{\alpha \mu \nu}^{{\rm F}}(q,k_1,k_2) &=& 
(k_1-k_2)_{\alpha} g_{\mu\nu} + 2q_{\nu}g_{\alpha\mu} 
- 2q_{\mu}g_{\alpha\nu}\,,  \nonumber\\
\Gamma_{\alpha \mu \nu}^{{\rm P}}(q,k_1,k_2) &=& 
 k_{2\nu} g_{\alpha\mu} - k_{1\mu}g_{\alpha\nu} \,.  
\label{GFGP}
\eea
The vertex $\Gamma_{\alpha \mu \nu}^{{\rm F}}(q,k_1,k_2)$ 
is Bose-symmetric only with respect to the
$\mu$ and $\nu$ legs. 
Evidently the above decomposition assigns a special role 
to the $q$-leg,
and allows $\Gamma_{\alpha \mu \nu}^{{\rm F}}(q,k_1,k_2)$ to satisfy the WI
\be 
q^{\alpha} \Gamma_{\alpha \mu \nu}^{{\rm F}}(q,k_2,k_1) = 
(k_2^2 - k_1^2)g_{\mu\nu} \,.
\label{WI2B}
\ee
where the right-hand side is the difference of two inverse 
propagators in the Feynman gauge \cite{CTxi}.
The term $\Gamma_{\alpha \mu \nu}^{{\rm P}}(q,k_1,k_2)$, which 
in configuration space corresponds to a pure divergence, 
contains the pinching momenta; as we will see in a moment,     
these momenta  
act on  the amplitude $A^a_{\mu} A^b_{\nu} \, q \,\bar{q}$ 
and trigger its STI.

In what  follows we  will carry  out the  analysis 
starting directly from the 
renormalizable (linear) Feynman gauge (RFG), {\it i.e.}  $\xi=1$; 
this does not constitute 
a loss of generality, provided that one is 
studying the entire $S$-matrix, as we do \cite{ASM}.   
This choice eliminates  the longitudinal momenta
from the tree-level propagators in Eq.(\ref{GluProp}),  
and allows us to focus our attention on the
all-order  study  of the effects of the 
longitudinal  momenta contained in 
$\Gamma_{\alpha \mu \nu}^{{\rm P}}(q,k_1,k_2)$. 

%%%%%%%%%%%%%%%%%%%%%%%%%%%%%%%%%%%%%%%%%%%%%%%%%%%%%%%%%%%%%%%%%%%%%%%
\begin{figure}[t]
\bce
\includegraphics[width=12cm]{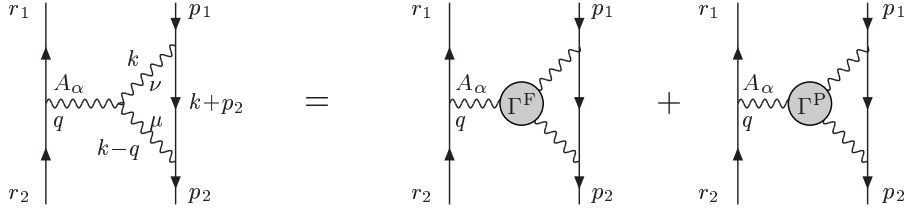}
\ece
\caption{\label{figA} Carrying out the fundamental PT vertex
  decomposition inside the tree-level three-gluon vertex.}
\end{figure}

\begin{figure}[b]
\bce
\includegraphics[width=6.5cm]{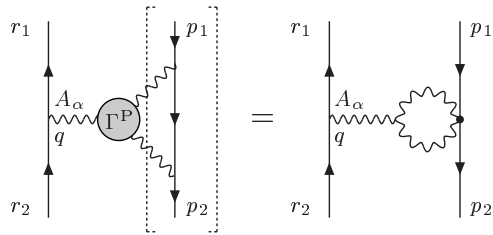}
\ece
\caption{\label{figB} The self-energy-like contribution coming from
  the pinching part of the tree-level three gluon vertex.}
\end{figure}

In order to appreciate the relevance of the 
amplitude $A^a_{\mu} A^b_{\nu} \, q \,\bar{q}$, let us remember the 
basic steps of the PT construction at one-loop.
To begin with, in the RFG 
the box is completely inert, since there are no  
pinching momenta, and therefore the PT box coincides with the 
conventional one computed at $\xi=1$. 
Then, in the non-Abelian vertex graph of Fig.\ref{figA} we 
carry out the splitting of the elementary three-gluon vertex 
given in Eq.(\ref{decomp}) (now $k_1 = k-q$ and $k_2 = -k$). 
Despite appearances, the part of the vertex graph containing
$\Gamma_{\alpha \mu \nu}^{{\rm P}}(q,k-q,-k)$ is in fact a 
propagator-like contribution: the longitudinal momenta 
of $\Gamma_{\alpha \mu \nu}^{{\rm P}}(q,k-q,-k)$ trigger 
the elementary WI  
$ k_{\nu}\gamma^{\nu} =  S^{-1}_0(\ksm + \psm)-  \,S^{-1}_0(\psm)$, 
whose first term removes (pinches out) 
 the internal bare fermion propagator $S_0(\ksm + \psm)$ 
(see Fig.\ref{figB}), 
whereas the second term dies on shell.
On the other hand, the part of the vertex graph containing 
$\Gamma_{\alpha \mu \nu}^{{\rm F}}(q,k-q,-k)$ remains unchanged, 
since it contains no longitudinal momenta; adding it to the usual 
Abelian-like graph (not shown) we obtain the one-loop 
PT vertex $\widehat\Gamma^{e\,[1]}_{\alpha}(q)$.
\begin{figure}[!t]
\bce
\includegraphics[width=15cm]{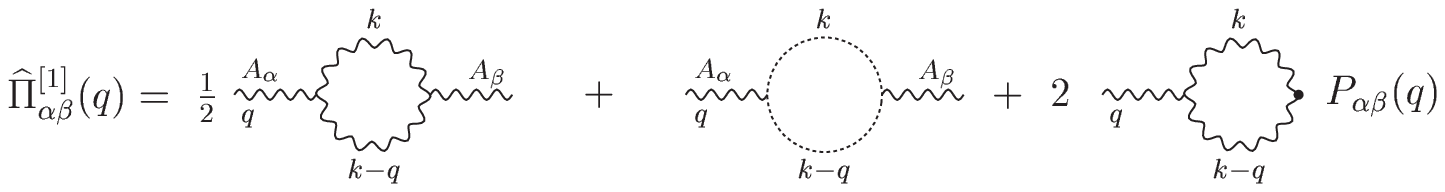}
\ece
\caption{\label{figG} Diagrammatic representation of the one-loop PT gluon
  self-energy $\widehat\Pi^{[1]}_{\alpha\beta}(q)$ as the sum of the
  conventional gluon self-energy and the pinch contributions coming
  from the vertex.}
\end{figure}
\begin{figure}[b]
\bce
\includegraphics[width=13cm]{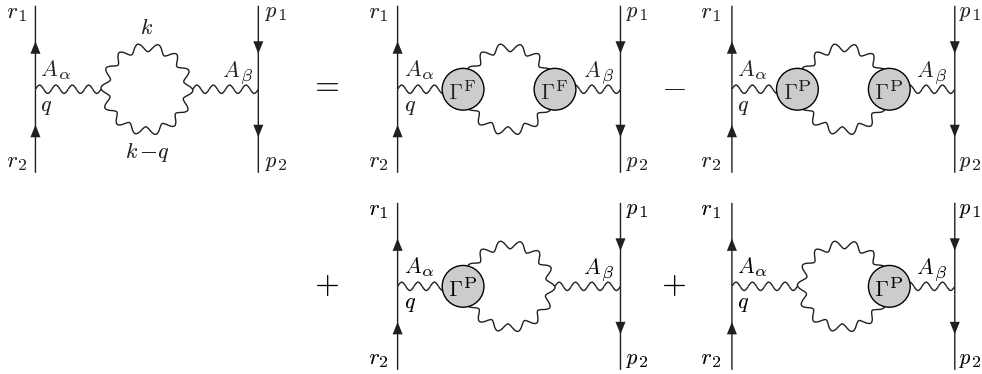}
\ece
\caption{\label{figC} The standard PT rearrangement of the two
  tree-level three-gluon vertices appearing in the self-energy diagram.}
\end{figure}

\begin{figure}[t]
\bce
\includegraphics[width=9cm]{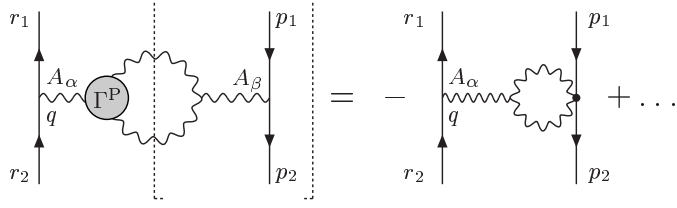}
\ece
\caption{\label{figD} The pinching term coming from the gluon
  self-energy diagram is minus the one that is extracted from the
  vertex diagrams.}
\end{figure}

The propagator-like parts extracted from the vertex are subsequently 
reassigned to the conventional self-energy graphs, giving rise 
to the one-loop PT gluon self-energy  
 $\widehat\Pi^{[1]}_{\alpha\beta}$ (Fig.\ref{figG}). Even though the answer is 
already contained in this sum, it is conceptually advantageous 
to trace down in more detail the exact fate of the pinch part.   
It turns out that this part cancels exactly against a corresponding 
term contained in the conventional self-energy graph.
To expose this cancellation, one carries out the following standard 
rearrangement of the two elementary three-gluon vertices
appearing in Fig.\ref{figC}:
\bea
\Gamma^{[0]}_{\alpha \mu \nu}\Gamma^{[0]\,\mu \nu}_{\beta}  &=&
[\Gamma^{{\rm F}}_{\alpha \mu \nu} 
+ \Gamma^{{\rm P}}_{\alpha \mu \nu}]
[\Gamma^{{\rm F}\,\mu \nu}_{\beta} +
\Gamma^{{\rm P}\,\mu\nu}_{\beta}]
\nonumber\\
&=&
\Gamma^{{\rm F}}_{\alpha \mu \nu}
\Gamma^{{\rm F}\,\mu \nu}_{\beta}- 
\Gamma^{{\rm P}}_{\beta \mu \nu}
\Gamma^{{\rm P}\,\mu\nu}_{\beta}+
\bigg\{\Gamma^{{\rm P}}_{\alpha \mu \nu}
\Gamma^{[0]\,\mu \nu}_{\beta}+
\Gamma^{[0]}_{\alpha \mu \nu}
\Gamma^{{\rm P}\,\mu \nu}_{\beta}\bigg\}.
\label{INPTDEC1}
\eea
This particular splitting, usually associated with the   
``intrinsic'' PT, allows for the identification of the term which will 
actually cancel against the pinch part coming from the vertex.
All one needs to do is recognize that the terms 
of Eq.(\ref{INPTDEC1}) appearing in curly brackets  
trigger the elementary WI 
\be 
k^{\nu} \Gamma_{\alpha \mu \nu}(q,k-q,-k) = 
\bigg[q^2 g_{\alpha\nu} - q_{\alpha}q_{\nu}\bigg] -
\bigg[\, (k-q)^2 g_{\alpha\nu} - (k-q)_{\alpha}(k-q)_{\nu}\,\bigg] \,
\label{WIN}
\ee
together with its Bose-symmetric counter-part  
from $(k-q)^{\mu} \Gamma_{\alpha \mu \nu}(q,k-q,-k)$.
Then it is elementary 
to verify that the term on the 
right-hand side proportional to  $[q^2 g_{\alpha\nu} - q_{\alpha}q_{\nu}]$
is the desired one (see Fig.\ref{figD}); 
incidentally, this is how the ``intrinsic'' PT works: one simply 
strips out all such terms from the conventional self-energy 
(first paper in \cite{Cornwall:1989gv};
see section V for more details).  
It must be clear from the above discussion that 
the PT rearrangement  
of terms between vertex- and self-energy graphs  
is actually encoded in the two graphs of Figs.\ref{figA} and \ref{figC}. 
Both graphs have    
the term $\Gamma_{\alpha \mu \nu}^{{\rm P}}(q,k-q,-k)$ common, 
whereas their terms in dotted brackets are the tree-level 
$t$-channel and $s$-channel contributions, respectively, to the 
four-particle amplitude $A^a_{\mu} A^b_{\nu} \, q \,\bar{q}$. 
As we will see in what follows, dressing the above amplitude 
with higher order corrections, 
and exploiting its STI, will  
provide us with the way of generalizing the PT to all orders.

With this intention in mind, 
of all the diagrams contributing to the QCD amplitude under consideration 
we will focus on  
the  subset of those graphs  which will receive  the action of
the     longitudinal    momenta     stemming     from    $\Gamma^{{\rm
P}}_{\alpha\mu\nu}(q,k_1,k_2)$, to be denoted by 
${\cal A}^{mnij}(r_1,r_2,p_1,p_2)$. It is given by 
\be 
{\cal A}^{mnij}(r_1,r_2,p_1,p_2)
= ig^2\bar  u^m(r_1) 
\frac{(\lambda^e)_{mn}}{2} \gamma_{\alpha} v^n(r_2) 
f^{eab} \Gamma^{\alpha\mu\nu}(q,k_1,k_2)
{\cal T}^{abij}_{\mu\nu}(k_1,k_2,p_1,p_2),
\ee
where $m,n,i,j=1,\dots,N$, are fundamental $SU(N)$ indices, 
$\lambda^e$   are   the
Gell-Mann   matrices,   and   ${\cal 
T}_{\mu\nu}^{abij}$     
is    the     sub-amplitude    $A_{\mu}^{a}(k_1)
A_{\nu}^{b}(k_2)\to q^i(p_1)\bar q^j(p_2)$, with the gluons {\it off-shell}
and    the   fermions    on-shell;    for   the    latter   
\be
\left.\bar
v(p_2)S^{-1}(p_2)\right|_{\psmp_2=m}=
\left.S^{-1}(p_1)u(p_1)\right|_{\psmp_1=m}=0,
\ee 
where  $S(p)$  is  the
(full) quark propagator, related to the corresponding quark
self-energy $\Sigma(p)$ through
\be
S(p)=\frac{i}{\psm-m-i\Sigma(p)}. 
\ee
Diagrammatically we have
\bce
\bpi(0,60)(0,-30)

\Text(-47,-7)[r]{${\cal A}^{mnij}=$}
\Text(-37,20)[l]{$\scriptstyle{m}$}
\Text(-37,-34)[l]{$\scriptstyle{n}$}
\ArrowLine(-15,-7)(-42,20)

\ArrowLine(-42,-34)(-15,-7)

\Text(-45,20)[r]{$\scriptstyle{r_1}$}
\Text(-45,-34)[r]{$\scriptstyle{r_2}$}
\Photon(-15,-7)(1,-7){1.5}{3}
\Text(-15,-13)[l]{$\scriptstyle{\alpha}$}
\Text(-15,-1)[l]{$\scriptstyle{e}$}

\PhotonArc(17.5,-7.5)(15,0,360){1.5}{17}
\GCirc(17.5,6.5){5}{0.8}
\Text(17.5,6.5)[c]{$\scriptstyle{\Delta}$}
\Text(5,7.5)[c]{$\scriptstyle{\nu}$}
\Text(32.5,7.5)[c]{$\scriptstyle{\sigma}$}
\GCirc(17.5,-22.5){5}{0.8}
\Text(17.5,-22.5)[c]{$\scriptstyle{\Delta}$}
\Text(5,-22.5)[c]{$\scriptstyle{\mu}$}
\Text(32.5,-22.5)[c]{$\scriptstyle{\rho}$}

\ArrowLine(65,20)(38,-7)
\Text(70,20)[l]{$\scriptstyle{p_1}$}
\Text(57.5,20)[r]{$\scriptstyle{i}$}

\ArrowLine(38,-7)(65,-34)
\Text(70,-34)[l]{$\scriptstyle{p_2}$}
\Text(57.5,-34)[r]{$\scriptstyle{j}$}

\GCirc(37.5,-7.5){10}{0.8}
\Text(37.5,-7.5)[c]{${\cal C}_{\rho\sigma}$} 

\DashLine(8,-39)(8,25){1}
\DashLine(8,25)(11,25){1}
\DashLine(8,-39)(11,-39){1}
\DashLine(80,-39)(80,25){1}
\DashLine(80,25)(77,25){1}
\DashLine(80,-39)(77,-39){1}

\epi
\ece
so that in terms of Green's functions the amplitude in brackets can be 
written as
\be
{\cal T}_{\mu\nu}^{abij} = 
\bar  v(p_2)\!\left[{\cal C}^{abij}_{\rho\sigma}(k_1,k_2,p_1,p_2)
\Delta^{\rho}_{\mu}(k_1)\Delta^{\sigma}_{\nu}(k_2)\right]\!u(p_1).
\label{sgf}
\ee
We next carry out the vertex decomposition of Eq.(\ref{decomp}), {\it
  i.e.} we write
\bce
\vspace{0.3cm}
\includegraphics[width=13cm]{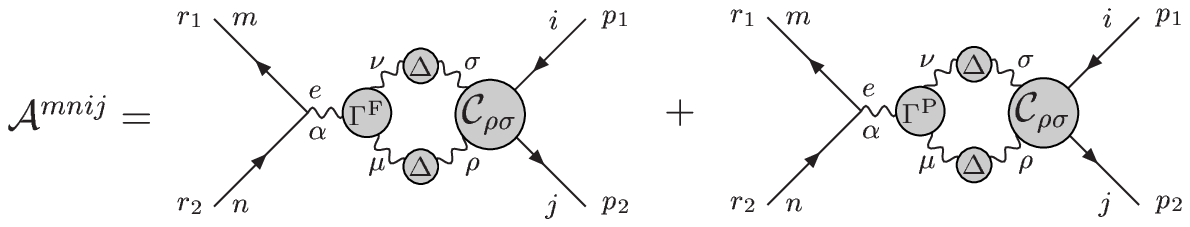}
\ece
Clearly,  there  is  an  equal contribution  from  the  $\Gamma^{{\rm
P}}$ situated on the right hand-side of the ${\cal
T}^{abij}_{\mu\nu}$ amplitude.

Let us then focus  on the STIs satisfied by the amplitude of Eq.(\ref{sgf}). 
For deriving them, we start from the following identities \cite{PasTar}
\bea
\bigg\langle T\left[\bar c^a(x)A^b_\nu(y)q^i(z)\bar q^j(w)
\right]\bigg\rangle&=&0 \,, \nonumber \\
\bigg\langle T\left[A^a_\mu(x)\bar c^b(y)q^i(z)\bar  q^j(w)
\right]\bigg\rangle&=&0 
\,,
\label{start}
\eea
valid due to ghost-charge conservation.
We then apply to the above equations the BRST operator $s$, which acts
on the fields as follows (recall that we work in the RFG,
{\it i.e.}, $\xi=1$)
\bea
s A^a_\mu(x)&=&\partial_\mu c^a(x)+gf^{acd}A^c_\mu(x)c^d(x), \nonumber
\\
s \bar c^a(x)&=&\partial^\mu A^a_\mu(x), \nonumber \\
s q^i(x)&=&ig\left[T^d\right]^{ik}c^d(x)q^k(x), 
\nonumber \\
s \bar q^i(x)&=&-ig\bar
q^k(x)\left[T^d\right]^{ki}c^d(x),
\label{BRST}
\eea
where $T^d$ are the $SU(N)$ generators. From Eq.(\ref{start}) we then find 
the identities
\bea
\partial_x^\mu C_{\mu\nu}^{abij}+\partial_\nu^yG_1^{abij}
+gf^{bcd} Q_{1\nu}^{acdij}+igX_{1\nu}^{abij}-
ig\bar X_{1\nu}^{abij}&=&0, \nonumber \\
\partial_y^\nu C_{\mu\nu}^{abij}+\partial_\mu^xG_2^{abij}
+gf^{acd} Q_{2\mu}^{cdbij}+igX_{2\mu}^{abij}
-ig\bar X_{2\mu}^{abij}&=&0,
\eea
where we have introduced the following
Green's functions (in configuration space)
\bea
C_{\mu\nu}^{abij}(x,y,z,w)&=&
\bigg\langle T\left[A^a_\mu(x)A^b_\nu(y)q^i(z)\bar q^j(w)
\right]\bigg\rangle, \nonumber \\
Q^{acdij}_{1\nu}(x,y,z,w)&=&
\bigg\langle T\left[\bar c^a(x)A^c_\nu(y)c^d(y)q^i(z)\bar q^j(w)
\right]\bigg\rangle, \nonumber \\
Q_{2\mu}^{cdbij}(x,y,z,w) &=&
\bigg\langle T\left[A^c_\mu(x)c^d(x)\bar c^b(y)q^i(z)\bar q^j(w)
\right]\bigg\rangle, \nonumber \\
G_1^{abij}(x,y,z,w) &=&\bigg\langle T\left[\bar
c^a(x)c^b(y)q^i(z)\bar q^j(w) \right]\bigg\rangle, \nonumber \\ 
G_2^{abij}(x,y,z,w) &=&\bigg\langle T\left[
c^a(x)\bar c^b(y)q^i(z)\bar q^j(w) \right]\bigg\rangle, \nonumber \\ 
X_{1\nu}^{abij}(x,y,z,w) &=&\bigg\langle T\left\{
\bar c^a(x) A^b_\nu(y)
\left[T^d\right]^{ik}c^d(z)q^k(z)\bar q^j(w)
\right\}\bigg\rangle, \nonumber \\
\bar X_{1\nu}^{abij}(x,y,z,w) &=&\bigg\langle T\left\{
\bar c^a(x) A^b_\nu(y)
q^i(z)\bar q^k(w)\left[T^d\right]^{kj}c^d(w)
\right\}\bigg\rangle, \nonumber \\
X_{2\mu}^{abij}(x,y,z,w) &=&\bigg\langle T\left\{
A^a_\mu(x)\bar c^b(y)
\left[T^d\right]^{ik}c^d(z)q^k(z)\bar q^j(w)
\right\}\bigg\rangle, \nonumber \\
\bar X_{2\mu}^{abij}(x,y,z,w) &=&\bigg\langle T\left\{
A^a_\mu(x)\bar c^b(y)
q^i(z)\bar q^k(w)\left[T^d\right]^{kj}c^d(w)
\right\}\bigg\rangle, 
\label{psgf}
\eea
After Fourier transform, the
above quantities define the following momentum-space Green's functions
\bea
C^{abij}_{\mu\nu}(k_1,k_2,p_1,p_2)&=&
{\cal C}^{abij}_{\rho\sigma}(k_1,k_2,p_1,p_2)
\Delta^{\rho}_{\mu}(k_1)\Delta^{\sigma}_{\nu}(k_2)S(p_1)S(p_2),
\nonumber \\
Q_{1\nu}^{acdij}(k_1,k_2,p_1,p_2)&=&{\cal
Q}_{1\nu}^{acdij}(k_1,k_2,p_1,p_2)D(k_1)S(p_1)S(p_2),\nonumber \\
Q_{2\mu}^{cdbij}(k_1,k_2,p_1,p_2)&=&{\cal
Q}_{2\mu}^{cdbij}(k_1,k_2,p_1,p_2)D(k_2)S(p_1)S(p_2),\nonumber \\
G_1^{abij}(k_1,k_2,p_1,p_2)&=&{\cal
G}_1^{abij}(k_1,k_2,p_1,p_2)D(k_1)D(k_2)S(p_1)S(p_2),\nonumber \\
G_2^{abij}(k_1,k_2,p_1,p_2)&=&{\cal
G}_2^{abij}(k_1,k_2,p_1,p_2)D(k_1)D(k_2)S(p_1)S(p_2),\nonumber \\
X_{1\nu}^{abij}(k_1,k_2,p_1,p_2)&=&{\cal
X}_{1\sigma}^{abij}(k_1,k_2,p_1,p_2)D(k_1)
\Delta^{\sigma}_{\nu}(k_2)S(p_2),\nonumber \\
\bar X_{1\nu}^{abij}(k_1,k_2,p_1,p_2)&=&\bar {\cal
X}_{1\sigma}^{abij}(k_1,k_2,p_1,p_2)D(k_1)
\Delta^{\sigma}_{\nu}(k_2)S(p_1),\nonumber \\
X_{2\mu}^{abij}(k_1,k_2,p_1,p_2)&=&{\cal
X}_{2\rho}^{abij}(k_1,k_2,p_1,p_2)
\Delta^{\rho}_{\mu}(k_1)D(k_2)S(p_2),\nonumber \\
\bar X_{2\mu}^{abij}(k_1,k_2,p_1,p_2)&=&\bar {\cal
X}_{2\rho}^{abij}(k_1,k_2,p_1,p_2)
\Delta^{\rho}_{\mu}(k_1)D(k_2)S(p_1).
\label{msgf}
\eea
In the above formulas,
all the momenta are supposed to be entering, {\it i.e.}, we have
\mbox{$k_1+k_2+p_1+p_2=0$}; moreover we have denoted by $D(k)$ and $\Delta_{\mu\nu}(k)$ 
the (full) RFG ghost and gluon 
propagators 
which are related to the corresponding 
ghost and gluon self-energies $L(k)$ and $\Pi_{\mu\nu}(k)$ through 
\bea
D(k)&=&\frac i{k^2-iL(k)}, \nonumber \\
\Delta_{\mu\nu}(k)&=&-i\left[\Delta(k^2)P_{\mu\nu}(k)+\frac{k_\mu
k_\nu}{k^4}\right], \qquad \Delta(k^2)=\frac 1{k^2+i\Pi(k^2)},
\eea
where $P_{\mu\nu}(k)=g_{\mu\nu} - k_\mu k_\nu/k^2$ represents the 
dimensionless projection operator, and
\mbox{$\Pi_{\mu\nu}(k)=\Pi(k^2)P_{\mu\nu}(k)$}.

Then, the above Green's functions have the following 
diagrammatic representation
%
%C
%
\bce
\bpi(0,50)(30,-20)

\Text(-15,-7.5)[r]{${C_{\mu\nu}}=$}

\Photon(0,20)(30,0){-1.5}{6}
\GCirc(16.5,8.5){5}{0.8}
\Text(16.5,8.5)[c]{$\scriptstyle{\Delta}$}
\Text(-5,20)[r]{$\scriptstyle{k_2}$}
\Text(9.5,21)[c]{$\scriptstyle{\nu}$}
\Text(29.5,8)[c]{$\scriptstyle{\sigma}$}

\Photon(0,-35)(30,-15){1.5}{6}
\GCirc(16.5,-23.5){5}{0.8}
\Text(16.5,-23.5)[c]{$\scriptstyle{\Delta}$}
\Text(-5,-35)[r]{$\scriptstyle{k_1}$}
\Text(9.5,-36)[c]{$\scriptstyle{\mu}$}
\Text(29.5,-23)[c]{$\scriptstyle{\rho}$}

\ArrowLine(60,10)(40,-5)
\ArrowLine(75,20)(60,10)
\GCirc(58.5,8.5){5}{0.8}
\Text(58.5,8.5)[c]{$\scriptstyle{S}$}
\Text(80,20)[l]{$\scriptstyle{p_1}$}

\ArrowLine(40,-10)(60,-25)
\ArrowLine(60,-25)(75,-35)
\GCirc(58.5,-23.5){5}{0.8}
\Text(58.5,-23.5)[c]{$\scriptstyle{S}$}
\Text(80,-35)[l]{$\scriptstyle{p_2}$}

\GCirc(37.5,-7.5){10}{0.8}
\Text(37.5,-7.5)[c]{${{\cal C}_{\rho\sigma}}$}

\epi
\ece
%
%Q1
%
\bce
\bpi(0,50)(110,0)

\Text(-20,-7.5)[r]{$Q_{1\nu}=$}

\DashArrowLine(-10,-7.5)(7.5,-7.5){1}
\DashArrowLine(11.5,-7.5)(30,-7.5){1}
\GCirc(10,-7.5){5}{0.8}
\Text(10,-7.5)[c]{$\scriptstyle{D}$}
\Text(-10,-15)[l]{$\scriptstyle{k_1}$}

\ArrowLine(60,10)(40,-5)
\ArrowLine(75,20)(60,10)
\GCirc(58.5,8.5){5}{0.8}
\Text(58.5,8.5)[c]{$\scriptstyle{S}$}
\Text(80,20)[l]{$\scriptstyle{p_1}$}

\ArrowLine(40,-10)(60,-25)
\ArrowLine(60,-25)(75,-35)
\GCirc(58.5,-23.5){5}{0.8}
\Text(58.5,-23.5)[c]{$\scriptstyle{S}$}
\Text(80,-35)[l]{$\scriptstyle{p_2}$}

\GCirc(37.5,-7.5){10}{0.8}
\Text(37.5,-7.5)[c]{${{\cal Q}_{1\nu}}$}

\Text(185,-7.5)[r]{${\cal Q}_{1\nu}=$}

\PhotonArc(217.5,-7.5)(15,0,180){1.5}{8.5}
\GCirc(217.5,6.5){5}{0.8}
\Text(218.5,6.5)[c]{$\scriptstyle{\Delta}$}
\Text(205,7.5)[c]{$\scriptstyle{\nu}$}
\Text(232.5,7.5)[c]{$\scriptstyle{\sigma}$}

\DashCArc(217.5,-7.5)(15,180,0){1}
\DashArrowLine(207,-18.1)(206.9,-18){1}
\DashArrowLine(228.1,-18.1)(228,-18.2){1}
\GCirc(217.5,-22.5){5}{0.8}
\Text(218.5,-22.5)[c]{$\scriptstyle{D}$}

\Vertex(202.5,-7.5){1.8}

\ArrowLine(265,15)(245,0)
\Text(270,15)[l]{$\scriptstyle{p_1}$}

\ArrowLine(240,-7.5)(275,-7.5)
\Text(280,-7.5)[l]{$\scriptstyle{p_2}$}

\DashArrowLine(265,-30)(245,-15){1}
\Text(270,-30)[l]{$\scriptstyle{k_1}$}

\GCirc(237.5,-7.5){10}{0.8}
\Text(238.5,-7.5)[c]{${\cal F}_{1\sigma}$}

\epi
\ece
%
%Q2
%
\bce
\bpi(0,70)(110,0)

\Text(-20,-7.5)[r]{$Q_{2\mu}=$}

\DashArrowLine(-10,-7.5)(7.5,-7.5){1}
\DashArrowLine(11.5,-7.5)(30,-7.5){1}
\GCirc(10,-7.5){5}{0.8}
\Text(10,-7.5)[c]{$\scriptstyle{D}$}
\Text(-10,-15)[l]{$\scriptstyle{k_2}$}

\ArrowLine(60,10)(40,-5)
\ArrowLine(75,20)(60,10)
\GCirc(58.5,8.5){5}{0.8}
\Text(58.5,8.5)[c]{$\scriptstyle{S}$}
\Text(80,20)[l]{$\scriptstyle{p_1}$}

\ArrowLine(40,-10)(60,-25)
\ArrowLine(60,-25)(75,-35)
\GCirc(58.5,-23.5){5}{0.8}
\Text(58.5,-23.5)[c]{$\scriptstyle{S}$}
\Text(80,-35)[l]{$\scriptstyle{p_2}$}

\GCirc(37.5,-7.5){10}{0.8}
\Text(37.5,-7.5)[c]{${{\cal Q}_{2\mu}}$}

\Text(185,-7.5)[r]{${\cal Q}_{2\mu}=$}

\PhotonArc(217.5,-7.5)(15,180,360){1.5}{8.5}

\Text(205,-22.5)[r]{$\scriptstyle{\mu}$}
\Text(232.5,-22.5)[c]{$\scriptstyle{\rho}$}

\DashCArc(217.5,-7.5)(15,0,180){1}
\DashArrowLine(207,3.1)(206.9,3){1}
\DashArrowLine(228.1,3.1)(228,3.2){1}
\GCirc(217.5,-22.5){5}{0.8}
\Text(218.5,-22.5)[c]{$\scriptstyle{\Delta}$}
\GCirc(217.5,6.5){5}{0.8}
\Text(218,6.5)[c]{$\scriptstyle{D}$}

\Vertex(202.5,-7.5){1.8}

\DashArrowLine(265,15)(245,0){1}
\Text(270,15)[l]{$\scriptstyle{k_2}$}

\ArrowLine(275,-7.5)(240,-7.5)
\Text(280,-7.5)[l]{$\scriptstyle{p_1}$}

\ArrowLine(245,-15)(265,-30)
\Text(270,-30)[l]{$\scriptstyle{p_2}$}

\GCirc(237.5,-7.5){10}{0.8}
\Text(238.5,-7.5)[c]{${\cal F}_{2\rho}$}

\epi
\ece
%
%G1, G2
%
\bce
\bpi(0,70)(115,0)

\Text(-15,-7.5)[r]{$G_1=$}

\DashArrowLine(13.5,10.5)(0,20){1}
\DashArrowLine(30,0)(21,5.5){1}
\GCirc(16.5,8.5){5}{0.8}
\Text(16.5,8.5)[c]{$\scriptstyle{D}$}
\Text(-5,20)[r]{$\scriptstyle{k_2}$}

\DashArrowLine(0,-35)(13.5,-25.5){1}
\DashArrowLine(21,-20.5)(30,-15){1}
\GCirc(16.5,-23.5){5}{0.8}
\Text(16.5,-23.5)[c]{$\scriptstyle{D}$}
\Text(-5,-35)[r]{$\scriptstyle{k_1}$}

\ArrowLine(60,10)(40,-5)
\ArrowLine(75,20)(60,10)
\GCirc(58.5,8.5){5}{0.8}
\Text(58.5,8.5)[c]{$\scriptstyle{S}$}
\Text(80,20)[l]{$\scriptstyle{p_1}$}

\ArrowLine(40,-10)(60,-25)
\ArrowLine(60,-25)(75,-35)
\GCirc(58.5,-23.5){5}{0.8}
\Text(58.5,-23.5)[c]{$\scriptstyle{S}$}
\Text(80,-35)[l]{$\scriptstyle{p_2}$}

\GCirc(37.5,-7.5){10}{0.8}
\Text(37.5,-7.5)[c]{${{\cal G}_1}$}

\Text(185,-7.5)[r]{$G_2=$}

\DashArrowLine(200,20)(213.5,10.5){1}
\DashArrowLine(221,5.5)(230,0){1}
\GCirc(216.5,8.5){5}{0.8}
\Text(216.5,8.5)[c]{$\scriptstyle{D}$}
\Text(195,20)[r]{$\scriptstyle{k_2}$}

\DashArrowLine(213.5,-25.5)(200,-35){1}
\DashArrowLine(230,-15)(221,-20.5){1}
\GCirc(216.5,-23.5){5}{0.8}
\Text(216.5,-23.5)[c]{$\scriptstyle{D}$}
\Text(195,-35)[r]{$\scriptstyle{k_1}$}

\ArrowLine(260,10)(240,-5)
\ArrowLine(275,20)(260,10)
\GCirc(258.5,8.5){5}{0.8}
\Text(258.5,8.5)[c]{$\scriptstyle{S}$}
\Text(280,20)[l]{$\scriptstyle{p_1}$}

\ArrowLine(240,-10)(260,-25)
\ArrowLine(260,-25)(275,-35)
\GCirc(258.5,-23.5){5}{0.8}
\Text(258.5,-23.5)[c]{$\scriptstyle{S}$}
\Text(280,-35)[l]{$\scriptstyle{p_2}$}

\GCirc(237.5,-7.5){10}{0.8}
\Text(237.5,-7.5)[c]{${{\cal G}_2}$}

\epi
\ece
%
%X1
%
\bce
\bpi(0,70)(115,0)

\Text(-20,-7.5)[r]{$X_{1\nu}=$}

\DashArrowLine(0,-35)(13.5,-25.5){1}
\DashArrowLine(21,-20.5)(30,-15){1}
\GCirc(16.5,-23.5){5}{0.8}
\Text(16.5,-23.5)[c]{$\scriptstyle{D}$}
\Text(-5,-35)[r]{$\scriptstyle{k_1}$}

\Photon(0,20)(30,0){-1.5}{6}
\GCirc(16.5,8.5){5}{0.8}
\Text(16.5,8.5)[c]{$\scriptstyle{\Delta}$}
\Text(-5,20)[r]{$\scriptstyle{k_2}$}
\Text(9.5,21)[c]{$\scriptstyle{\nu}$}
\Text(29.5,8)[c]{$\scriptstyle{\sigma}$}

\ArrowLine(45,-7.5)(62.5,-7.5)
\ArrowLine(67.5,-7.5)(82.5,-7.5)
\GCirc(65,-7.5){5}{0.8}
\Text(65,-7.5)[c]{$\scriptstyle{S}$}
\Text(90,-7.5)[c]{$\scriptstyle{p_2}$}

\GCirc(37.5,-7.5){10}{0.8}
\Text(37.5,-7.5)[c]{${{\cal X}_{1\sigma}}$}

\Text(185,-7.5)[r]{${\cal X}_{1\sigma}=$}

\DashCArc(257.5,-7.5)(15,180,0){1}
\GCirc(257.5,-22.5){5}{0.8}
\Text(258.5,-22.5)[c]{$\scriptstyle{D}$}
\DashArrowLine(268.1,-18.1)(268.2,-18){1}
\DashArrowLine(246.9,-18.1)(247,-18.2){1}

\CArc(257.5,-7.5)(15,0,180)
\GCirc(257.5,7.5){5}{0.8}
\Text(258.5,7.5)[c]{$\scriptstyle{S}$}
\DashArrowLine(268.2,3)(268.1,3.1){1}
\DashArrowLine(247,3.1)(246.9,3){1}

\Vertex(272.5,-7.5){1.8}

\Photon(210,15)(230,0){1.5}{5}
\Text(205,15)[r]{$\scriptstyle{k_2}$}

\DashArrowLine(200,-7.5)(235,-7.5){1}
\Text(195,-12.5)[l]{$\scriptstyle{k_1}$}

\ArrowLine(230,-15)(210,-30)
\Text(205,-30)[r]{$\scriptstyle{p_2}$}

\GCirc(237.5,-7.5){10}{0.8}
\Text(238.5,-7.5)[c]{${\cal I}_{1\sigma}$}

\epi
\ece
%
%X1 bar
%
\bce
\bpi(0,70)(115,0)

\Text(-20,-7.5)[r]{$\bar X_{1\nu}=$}

\DashArrowLine(0,-35)(13.5,-25.5){1}
\DashArrowLine(21,-20.5)(30,-15){1}
\GCirc(16.5,-23.5){5}{0.8}
\Text(16.5,-23.5)[c]{$\scriptstyle{D}$}
\Text(-5,-35)[r]{$\scriptstyle{k_1}$}

\Photon(0,20)(30,0){-1.5}{6}
\GCirc(16.5,8.5){5}{0.8}
\Text(16.5,8.5)[c]{$\scriptstyle{\Delta}$}
\Text(-5,20)[r]{$\scriptstyle{k_2}$}
\Text(9.5,21)[c]{$\scriptstyle{\nu}$}
\Text(29.5,8)[c]{$\scriptstyle{\sigma}$}

\ArrowLine(62.5,-7.5)(45,-7.5)
\ArrowLine(82.5,-7.5)(67.5,-7.5)
\GCirc(65,-7.5){5}{0.8}
\Text(65,-7.5)[c]{$\scriptstyle{S}$}
\Text(90,-7.5)[c]{$\scriptstyle{p_1}$}

\GCirc(37.5,-7.5){10}{0.8}
\Text(37.5,-7.5)[c]{${\bar{\cal X}_{1\sigma}}$}

\Text(185,-7.5)[r]{$\bar{\cal X}_{1\sigma}=$}

\CArc(257.5,-7.5)(15,180,0)
\GCirc(257.5,-22.5){5}{0.8}
\Text(258.5,-22.5)[c]{$\scriptstyle{S}$}
\DashArrowLine(268.1,-18.1)(268.2,-18){1}
\DashArrowLine(246.9,-18.1)(247,-18.2){1}

\DashCArc(257.5,-7.5)(15,0,180){1}
\GCirc(257.5,7.5){5}{0.8}
\Text(258.5,7.5)[c]{$\scriptstyle{D}$}
\DashArrowLine(268.1,3.1)(268.2,3){1}
\DashArrowLine(246.9,3)(247,3.1){1}

\Vertex(272.5,-7.5){1.8}

\ArrowLine(210,15)(230,0)
\Text(205,15)[r]{$\scriptstyle{p_1}$}

\Photon(235,-7.5)(200,-7.5){1.5}{5}
\Text(195,-12.5)[l]{$\scriptstyle{k_2}$}

\DashArrowLine(210,-30)(230,-15){1}
\Text(205,-30)[r]{$\scriptstyle{k_1}$}

\GCirc(237.5,-7.5){10}{0.8}
\Text(238.5,-7.5)[c]{$\bar{\cal I}_{1\sigma}$}

\epi
\ece
%
%X2
%
\bce
\bpi(0,70)(115,0)

\Text(-20,-7.5)[r]{$X_{2\mu}=$}

\DashArrowLine(0,20)(13.5,10.5){1}
\DashArrowLine(21,5.5)(30,0){1}
\GCirc(16.5,8.5){5}{0.8}
\Text(16.5,8.5)[c]{$\scriptstyle{D}$}
\Text(-5,20)[r]{$\scriptstyle{k_2}$}

\Photon(0,-35)(30,-15){1.5}{6}
\GCirc(16.5,-23.5){5}{0.8}
\Text(16.5,-23.5)[c]{$\scriptstyle{\Delta}$}
\Text(-5,-35)[r]{$\scriptstyle{k_1}$}
\Text(9.5,-36)[c]{$\scriptstyle{\mu}$}
\Text(29.5,-23)[c]{$\scriptstyle{\rho}$}

\ArrowLine(45,-7.5)(62.5,-7.5)
\ArrowLine(67.5,-7.5)(82.5,-7.5)
\GCirc(65,-7.5){5}{0.8}
\Text(65,-7.5)[c]{$\scriptstyle{S}$}
\Text(90,-7.5)[c]{$\scriptstyle{p_2}$}

\GCirc(37.5,-7.5){10}{0.8}
\Text(37.5,-7.5)[c]{${{\cal X}_{2\rho}}$}

\Text(185,-7.5)[r]{${\cal X}_{2\rho}=$}

\DashCArc(257.5,-7.5)(15,180,0){1}
\GCirc(257.5,-22.5){5}{0.8}
\Text(258.5,-22.5)[c]{$\scriptstyle{D}$}
\DashArrowLine(268.1,-18.1)(268.2,-18){1}
\DashArrowLine(246.9,-18.1)(247,-18.2){1}

\CArc(257.5,-7.5)(15,0,180)
\GCirc(257.5,7.5){5}{0.8}
\Text(258.5,7.5)[c]{$\scriptstyle{S}$}
\DashArrowLine(268.2,3)(268.1,3.1){1}
\DashArrowLine(247,3.1)(246.9,3){1}

\Vertex(272.5,-7.5){1.8}

\DashArrowLine(210,15)(230,0){1}
\Text(205,15)[r]{$\scriptstyle{k_2}$}

\Photon(235,-7.5)(200,-7.5){1.5}{5}
\Text(195,-12.5)[l]{$\scriptstyle{k_1}$}

\ArrowLine(230,-15)(210,-30)
\Text(205,-30)[r]{$\scriptstyle{p_2}$}

\GCirc(237.5,-7.5){10}{0.8}
\Text(238.5,-7.5)[c]{${\cal I}_{2\rho}$}

\epi
\ece
%
%X2 bar
%
\bce
\bpi(0,105)(115,-35)

\Text(-20,-7.5)[r]{$\bar X_{2\mu}=$}

\DashArrowLine(13.5,10.5)(0,20){1}
\DashArrowLine(30,0)(21,5.5){1}
\GCirc(16.5,8.5){5}{0.8}
\Text(16.5,8.5)[c]{$\scriptstyle{D}$}
\Text(-5,20)[r]{$\scriptstyle{k_2}$}

\Photon(0,-35)(30,-15){1.5}{6}
\GCirc(16.5,-23.5){5}{0.8}
\Text(16.5,-23.5)[c]{$\scriptstyle{\Delta}$}
\Text(-5,-35)[r]{$\scriptstyle{k_1}$}
\Text(9.5,-36)[c]{$\scriptstyle{\mu}$}
\Text(29.5,-23)[c]{$\scriptstyle{\rho}$}

\ArrowLine(62.5,-7.5)(45,-7.5)
\ArrowLine(82.5,-7.5)(67.5,-7.5)
\GCirc(65,-7.5){5}{0.8}
\Text(65,-7.5)[c]{$\scriptstyle{S}$}
\Text(90,-7.5)[c]{$\scriptstyle{p_1}$}

\GCirc(37.5,-7.5){10}{0.8}
\Text(37.5,-7.5)[c]{${\bar{\cal X}_{2\rho}}$}

\Text(185,-7.5)[r]{$\bar{\cal X}_{2\rho}=$}

\CArc(257.5,-7.5)(15,180,0)
\GCirc(257.5,-22.5){5}{0.8}
\Text(258.5,-22.5)[c]{$\scriptstyle{S}$}
\DashArrowLine(268.1,-18.1)(268.2,-18){1}
\DashArrowLine(246.9,-18.1)(247,-18.2){1}

\DashCArc(257.5,-7.5)(15,0,180){1}
\GCirc(257.5,7.5){5}{0.8}
\Text(258.5,7.5)[c]{$\scriptstyle{D}$}
\DashArrowLine(268.1,3.1)(268.2,3){1}
\DashArrowLine(246.9,3)(247,3.1){1}

\Vertex(272.5,-7.5){1.8}

\ArrowLine(210,15)(230,0)
\Text(205,15)[r]{$\scriptstyle{p_1}$}

\DashArrowLine(200,-7.5)(235,-7.5){1}
\Text(195,-12.5)[l]{$\scriptstyle{k_2}$}

\Photon(230,-15)(210,-30){1.5}{5}
\Text(205,-30)[r]{$\scriptstyle{k_1}$}

\GCirc(237.5,-7.5){10}{0.8}
\Text(238.5,-7.5)[c]{$\bar{\cal I}_{2\rho}$}

\epi
\ece

In terms of these quantities, the needed STIs read
\bea
k_1^\mu C_{\mu\nu}^{abij}+k_{2\nu}G_1^{abij}-igf^{bcd}
Q_{1\nu}^{acdij}+gX_{1\nu}^{abij}
-g\bar X_{1\nu}^{abij}&=&0, \nonumber \\
k_2^\nu C_{\mu\nu}^{abij}+k_{1\mu}G_2^{abij}-igf^{acd}
Q_{2\mu}^{cdbij}+gX_{2\mu}^{abij}
-g\bar X_{2\mu}^{abij}&=&0.
\label{BasSTI}
\eea  
As  can be clearly seen 
in   the above  diagrammatic  representation,   the  terms
$X_{1\nu}$,   $\bar  X_{1\nu}$,   $X_{2\mu}$   and  $\bar   X_{2\mu}$,
correspond  to terms  that die  on-shell, since  they are  missing one
fermion  propagator;  at  lowest   order  they  are  simply  the  terms
proportional  to  the  inverse  tree-level propagators  $(\psm_1+m)$  and
$(\psm_2-m)$ appearing  in the PT calculations. Indeed,  we multiply
both sides  of Eq.(\ref{BasSTI}) by  the product $S^{-1}(p_1)S^{-1}(p_2)$
of  the two  inverse propagators  of the  external fermions,  and then
sandwich the  resulting amplitude  between the on-shell  spinors $\bar
v(p_1)$ and $u(p_2)$. Since the fermion are assumed to be on-shell, by
virtue  of the  Dirac equation the vanishing of the aforementioned
terms follows. Thus we arrive at the on-shell STIs
\bea
k_1^\mu {\cal T}_{\mu\nu}^{abij}(k_1,k_2,p_1,p_2)&=&
{\cal S}^{abij}_{1\nu}(k_1,k_2,p_1,p_2), \nonumber \\
k_2^\nu {\cal T}_{\mu\nu}^{abij}(k_1,k_2,p_1,p_2)&=&
{\cal S}^{abij}_{2\mu}(k_1,k_2,p_1,p_2),
\label{onshSTI}
\eea
where
\bea
{\cal S}^{abij}_{1\nu}(k_1,k_2,p_1,p_2)&=&
\bar  v(p_1)\left[igf^{bcd}{\cal
Q}_{1\nu}^{acdij}(k_1,k_2,p_1,p_2)D(k_1)\right.\nonumber \\
&-&\left.
k_{2\nu}{\cal
G}_1^{abij}(k_1,k_2,p_1,p_2)D(k_1)D(k_2)\right]u(p_2), \nonumber \\
{\cal S}^{abij}_{2\mu}(k_1,k_2,p_1,p_2)&=&
\bar  v(p_1)\left[igf^{acd}{\cal
Q}_{2\mu}^{cdbij}(k_1,k_2,p_1,p_2)D(k_2)\right.\nonumber \\
&-&\left.
k_{1\mu}{\cal
G}_2^{abij}(k_1,k_2,p_1,p_2)D(k_1)D(k_2)\right]u(p_2).
\label{onshdef}
\eea
Perturbatively, the above equations are of the (schematic) form
\bea
{\cal T}^{[n]}&=&{\cal C}^{[n_1]}\Delta^{[n_2]}\Delta^{[n_3]},\nonumber \\
{\cal S}_1^{[n]}&=&{\cal Q}_1^{[m_1]}D^{[m_2]}-{\cal G}_1^{[\ell_1]}
D^{[\ell_2]}D^{[\ell_3]},\nonumber \\
{\cal S}_2^{[n]}&=&{\cal Q}_2^{[m_1]}D^{[m_2]}-{\cal G}_2^{[\ell_1]}
D^{[\ell_2]}D^{[\ell_3]},
\label{schemnot}
\eea
with $n_1+n_2+n_3=m_1+m_2=\ell_1+\ell_2+\ell_3=n$.

Since the external (on-shell) quarks are inert throughout 
our construction, and in order 
to avoid notational clutter, in what follows we will suppress both the
color indices $i$ and $j$ of the fundamental $SU(N)$ representation, 
and denote through the label $\pci$ ($i=1,2$) the dependence on the (on-shell) 
momenta $p_1$ and $p_2$.

\section{\label{sec:FC}The fundamental cancellations}

Having established the STIs of Eq.(\ref{onshSTI}) we now turn to
the main conceptual point related  to the all orders PT construction. 
The basic observation is
the   following.   In  perturbation   theory  the   quantities  ${\cal
T}_{\mu\nu}^{ab}$, ${\cal S}_{1\nu}^{ab}$, and
${\cal S}_{2\mu}^{ab}$ are given by Feynman  diagrams,
which can be separated
into  distinct classes,  depending on  their kinematic  dependence and
their geometrical properties.  Graphs which do not contain information
about  the  kinematical  details   of  the  incoming  test-quarks  are
self-energy graphs,  whereas those which  display a dependence  on the
test quarks are vertex graphs. The  former depend only on the variable
$s$,  whereas the latter  on both  $s$ and  the mass  $m$ of  the test
quarks; equivalently,  we  will  refer to  them  as $s$-channel  or
$t$-channel  graphs,   respectively. In  addition to this $s$/$t$
decomposition, the Feynman diagrams  allow for the distinction between
one-particle irreducible  (1PI)  and  one-particle  reducible  (1PR)
graphs. Thus, 1PR graphs are  those which, after cutting one line, get
disconnected into two subgraphs none of which is a tree-level graph; if
this does  not happen, then the graph  is 1PI.  The  distinction between 1PI
and  1PR  is  necessary  for  constructing  eventually  effective  1PI
$n$-point  functions.  Thus,  the Feynman  graphs allow  the following
separation
\bea
{\cal T}_{\mu\nu}^{ab}&=&[{\cal
T}_{\mu\nu}^{ab}]_{s,\scriptscriptstyle{\rm I}}+ 
[{\cal
T}_{\mu\nu}^{ab}]_{s,\scriptscriptstyle{\rm R}}+
[{\cal
T}_{\mu\nu}^{ab}]_{t,\scriptscriptstyle{\rm I}}+ 
[{\cal
T}_{\mu\nu}^{ab}]_{t,\scriptscriptstyle{\rm R}}, \nonumber\\
{\cal S}_{1\nu}^{ab}&=&[{\cal
S}_{1\nu}^{ab}]_{s,\scriptscriptstyle{\rm I}}+ 
[{\cal
S}_{1\nu}^{ab}]_{s,\scriptscriptstyle{\rm R}}+[{\cal
S}_{1\nu}^{ab}]_{t,\scriptscriptstyle{\rm I}}+ 
[{\cal
S}_{1\nu}^{ab}]_{t,\scriptscriptstyle{\rm R}},
\nonumber \\
{\cal S}_{2\mu}^{ab}&=&[{\cal
S}_{2\mu}^{ab}]_{s,\scriptscriptstyle{\rm I}}+ 
[{\cal
S}_{2\mu}^{ab}]_{s,\scriptscriptstyle{\rm R}}+[{\cal
S}_{2\mu}^{ab}]_{t,\scriptscriptstyle{\rm I}}+ 
[{\cal
S}_{2\mu}^{ab}]_{t,\scriptscriptstyle{\rm R}}.
\label{stsa}
\eea
For example at order $n$ one has the following decomposition
\bea
\nonumber \\
\left[{\cal T}^{[n]}_{\mu\nu}\right]_{s,\scriptscriptstyle{\rm I}}=\quad
\bpi(0,0)(0,-3)

\Photon(-5,25)(20,0){1.5}{5}
\Text(0,26)[l]{$\scriptstyle{\nu}$}
\Text(0,-26)[l]{$\scriptstyle{\mu}$}
\GCirc(5.5,14.5){5}{0.8}
\Text(5.5,14.5)[c]{$\scriptstyle{\Delta}$}
\Photon(-5,-25)(20,0){1.5}{5}
\GCirc(5.5,-14.5){5}{0.8}
\Text(5.5,-14.5)[c]{$\scriptstyle{\Delta}$}
\Photon(20,0)(50,0){1.5}{5}
\GCirc(20,0){7.5}{0.8}
\Text(20,0)[c]{\scriptsize{1PI}}
\Text(20,14)[c]{$\scriptstyle{[n]}$}
\ArrowLine(75,25)(50,0)
\ArrowLine(50,0)(75,-25)

\Text(90,0)[c]{$+$}

\Photon(105,25)(130,0){1.5}{5}
\Text(110,26)[l]{$\scriptstyle{\nu}$}
\Text(110,-26)[l]{$\scriptstyle{\mu}$}
\GCirc(115.5,14.5){5}{0.8}
\Text(116,14.5)[c]{$\scriptstyle{\Delta}$}
\Photon(105,-25)(130,0){-1.5}{5}
\GCirc(115.5,-14.5){5}{0.8}
\Text(116,-14.5)[c]{$\scriptstyle{\Delta}$}
\Photon(130,0)(160,0){1.5}{5}
\GCirc(145,0){7.5}{0.8}
\Text(145,0)[c]{\scriptsize{1PI}}
\Text(145,14)[c]{$\scriptstyle{[n]}$}
\ArrowLine(185,25)(160,0)
\ArrowLine(160,0)(185,-25)

\epi
\qquad\qquad\qquad&\qquad&\qquad\qquad\qquad\qquad\qquad
\nonumber \\
\nonumber \\
\nonumber \\
\left[{\cal
T}^{[n]}_{\mu\nu}\right]_{s,\scriptscriptstyle{\rm R}}=\quad
\bpi(0,0)(0,-3)

\Photon(-5,25)(20,0){1.5}{5}
\Text(0,26)[l]{$\scriptstyle{\nu}$}
\Text(0,-26)[l]{$\scriptstyle{\mu}$}
\GCirc(5.5,14.5){5}{0.8}
\Text(5.5,14.5)[c]{$\scriptstyle{\Delta}$}
\Photon(-5,-25)(20,0){1.5}{5}
\GCirc(5.5,-14.5){5}{0.8}
\Text(5.5,-14.5)[c]{$\scriptstyle{\Delta}$}
\Photon(20,0)(60,0){1.5}{6}
\GCirc(20,0){7.5}{0.8}
\Text(20,0)[c]{\scriptsize{1PI}}
\Text(20,14)[c]{$\scriptstyle{[n_1]}$}
\ArrowLine(85,25)(60,0)
\ArrowLine(60,0)(85,-25)
\GCirc(45,0){7.5}{0.8}
\Text(45,0)[c]{$\scriptstyle{\Delta}$}
\Text(45,14)[c]{$\scriptstyle{[n_2]}$}

\Text(110,0)[l]{$n_1+n_2=n,\ n_1<n$}

\epi
\qquad\qquad\qquad&\qquad&
\nonumber \\
\nonumber \\
\nonumber \\
\left[{\cal
T}^{[n]}_{\mu\nu}\right]_{t,\scriptscriptstyle{\rm I}}=\quad
\bpi(0,0)(0,-3)

\Photon(-5,25)(20,0){1.5}{5}
\Text(0,26)[l]{$\scriptstyle{\nu}$}
\Text(0,-26)[l]{$\scriptstyle{\mu}$}
\GCirc(5.5,14.5){5}{0.8}
\Text(5.5,14.5)[c]{$\scriptstyle{\Delta}$}
\Photon(-5,-25)(20,0){1.5}{5}
\GCirc(5.5,-14.5){5}{0.8}
\Text(5.5,-14.5)[c]{$\scriptstyle{\Delta}$}
\ArrowLine(45,25)(20,0)
\ArrowLine(20,0)(45,-25)
\GCirc(20,0){7.5}{0.8}
\Text(20,0)[c]{\scriptsize{1PI}}
\Text(20,14)[c]{$\scriptstyle{[n]}$}

\epi
\qquad\qquad\qquad&\qquad&
\nonumber \\
\nonumber \\
\nonumber \\
\left[{\cal
T}_{\mu\nu}^{[n]}\right]_{t,\scriptscriptstyle{\rm R}}=\quad
\bpi(0,0)(0,-3)

\Photon(-5,25)(20,0){1.5}{5}
\Text(0,26)[l]{$\scriptstyle{\nu}$}
\Text(0,-26)[l]{$\scriptstyle{\mu}$}
\GCirc(5.5,14.5){5}{0.8}
\Text(5.5,14.5)[c]{$\scriptstyle{\Delta}$}
\Photon(-5,-25)(20,0){1.5}{5}
\GCirc(5.5,-14.5){5}{0.8}
\Text(5.5,-14.5)[c]{$\scriptstyle{\Delta}$}
\Photon(20,0)(70,0){1.5}{7}
\GCirc(20,0){7.5}{0.8}
\Text(20,0)[c]{\scriptsize{1PI}}
\Text(20,14)[c]{$\scriptstyle{[n_1]}$}
\GCirc(45,0){7.5}{0.8}
\Text(45,0)[c]{$\scriptstyle{\Delta}$}
\Text(45,14)[c]{$\scriptstyle{[n_2]}$}
\ArrowLine(95,25)(70,0)
\ArrowLine(70,0)(95,-25)
\GCirc(70,0){7.5}{0.8}
\Text(70,0)[c]{\scriptsize{1PI}}
\Text(70,14)[c]{$\scriptstyle{[n_3]}$}

\Text(110,0)[l]{$n_1+n_2+n_3=n,\ n_3>0$}

\epi
\qquad\qquad\qquad&\qquad& \\ \nonumber
\eea
(notice that when $n_2=n$ in $[{\cal
T}_{\mu\nu}^{[n]}]_{s,\scriptscriptstyle{\rm R}}$, one has to remove
from the full propagator $\Delta^{[n]}$ the 1PI part).

The crucial point is that the action of the momenta 
$k_1^\mu$ or $k_2^\nu$ on 
${\cal T}_{\mu\nu}^{ab}$ does not respect, in general,
the above separation. 
In fact, whereas the characterization of graphs as propagator-like
and vertex-like 
can be done unambiguously in the absence of
longitudinal momenta ({\it e.g.}, in a purely scalar theory), 
their presence mixes propagator- with vertex-like terms, since the
last ones generate (through pinching) effectively 
propagator-like terms.
Similarly, the separation between 1PI and 1PR terms is no longer
unambiguous, since  
1PR diagrams are 
converted (through pinching) into effectively 1PI ones
(the opposite cannot happen). In particular this last effect is
most easily seen in the so-called intrinsic PT construction (see Section
\ref{sec:IPT}), where the order $n$ 1PI self-energy diagrams receive
1PI contributions from the 1PR strings made of
self-energy insertions of order less than $n$ (incidentally notice that
these last contributions are in
fact instrumental for ensuring that in the Standard Model
the PT resummed propagator does not shift the position of the pole, as 
has been shown in the second paper of \cite{Papavassiliou:1995fq}).

\begin{figure}[b]
\bce
\includegraphics[width=16.4cm]{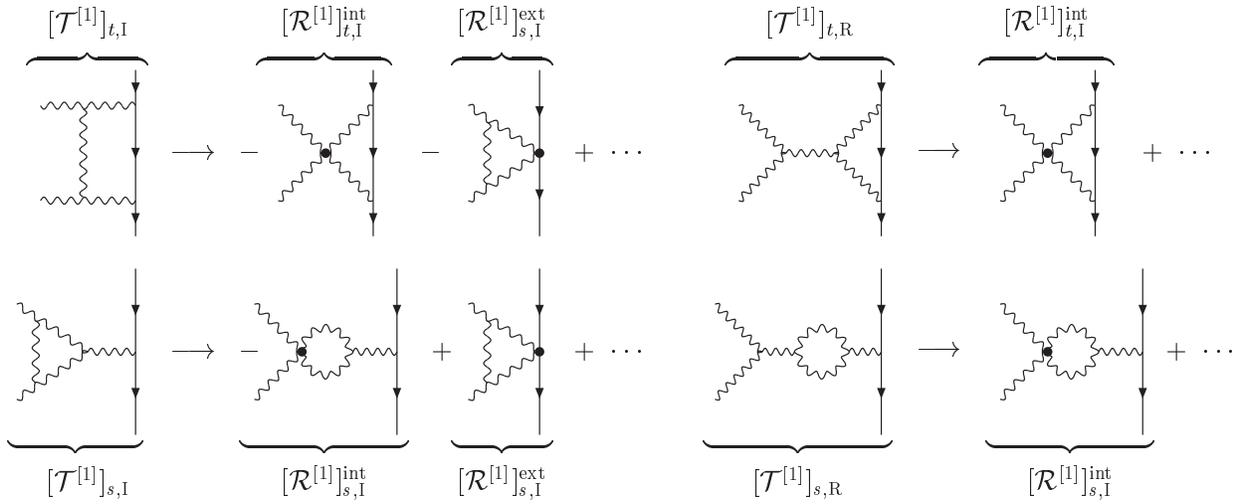}	
\ece
\caption{\label{figE} Some two-loop examples of the ${\cal R}$ terms,
  together with the Feynman diagrams from which they originate.}
\end{figure}

Then, even though Eq.(\ref{onshSTI}) holds for the entire amplitude 
${\cal T}_{\mu\nu}^{ab}$, 
it is not true for the individual sub-amplitudes
defined in Eq.(\ref{stsa}), {\it i.e.}, we have 
\bea
k^\mu_1 [{\cal T}^{ab}_{\mu\nu}]_{x,\scriptscriptstyle{\rm Y}} &\neq&  
[{\cal S}_{1\nu}^{ab}]_{x,\scriptscriptstyle{\rm
Y}},  \nonumber \\
k^\nu_2 [{\cal T}^{ab}_{\mu\nu}]_{x,\scriptscriptstyle{\rm Y}} &\neq&  
[{\cal S}_{2\mu}^{ab}]_{x,\scriptscriptstyle{\rm
Y}}, \qquad x=s,t;\ {\rm Y=I,R},
\label{neq}
\eea
where $I$ (respectively $R$) indicates the one-particle {\it
irreducible} (respectively {\it reducible}) parts of the amplitude
involved. The
reason  for  this  inequality, are  precisely  the
propagator-like  terms, such  as  those encountered  in  the one-  and
two-loop calculations (see Fig.\ref{figE}).

In particular, for individual sub-amplitudes we have that
\bea
k_1^\mu[{\cal T}^{ab}_{\mu\nu}]_{s,\scriptscriptstyle{\rm R}}
&=&[{\cal S}^{ab}_{1\nu}]_{s,\scriptscriptstyle{\rm R}} 
+[{\cal R}^{ab}_{1\nu}]_{s,\scriptscriptstyle{\rm
I}}^{\rm int},
\nonumber\\
k_1^\mu[{\cal T}^{ab}_{\mu\nu}]_{s,\scriptscriptstyle{\rm I}}
&=& [{\cal S}^{ab}_{1\nu}]_{s,\scriptscriptstyle{\rm I}} 
-[{\cal R}^{ab}_{1\nu}]_{s,\scriptscriptstyle{\rm I}}^{\rm int}
+[{\cal R}^{ab}_{1\nu}]_{s,\scriptscriptstyle{\rm I}}^{\rm ext}, \nonumber\\
k_1^\mu[{\cal T}^{ab}_{\mu\nu}]_{t,\scriptscriptstyle{\rm R}}
&=& [{\cal S}^{ab}_{1\nu}]_{t,\scriptscriptstyle{\rm R}} 
+[{\cal R}^{ab}_{1\nu}]_{t,\scriptscriptstyle{\rm I}}^{\rm int},
\nonumber\\
k_1^\mu[{\cal T}^{ab}_{\mu\nu}]_{t,\scriptscriptstyle{\rm I}}
&=& [{\cal S}^{ab}_{1\nu}]_{t,\scriptscriptstyle{\rm I}} 
-[{\cal R}^{ab}_{1\nu}]_{t,\scriptscriptstyle{\rm I}}^{\rm int}
-[{\cal R}^{ab}_{1\nu}]_{s,\scriptscriptstyle{\rm I}}^{\rm ext},
\label{xYres}
\eea
with similar equations holding when acting with the momentum $k_2^\nu$. In the
above equations the superscript ``ext'' and ``int'' stands for
``external'' and ``internal'' respectively, and refers to whether or not
the pinching part of the diagram at hands has touched the external
on-shell fermion legs. At order $n$, some example of the ${\cal R}^{[n]}$ terms
introduced in the equations above are the following
\bea
\nonumber \\
\big[{\cal R}^{[n]}_{1\nu}\big]_{s, {\rm I}}^{{\rm int}}\subset\quad
\bpi(0,0)(0,-3)

\Photon(-5,25)(20,0){1.5}{5}
\Text(0,26)[l]{$\scriptstyle{\nu}$}
\GCirc(5.5,14.5){5}{0.8}
\Text(5.5,14.5)[c]{$\scriptstyle{\Delta}$}
\Photon(-5,-25)(20,0){1.5}{5}
\Photon(20,0)(60,0){1.5}{6}
\GCirc(20,0){7.5}{0.8}
\Text(20,0)[c]{\scriptsize{1PI}}
\Text(20,14)[c]{$\scriptstyle{[n_1]}$}
\GCirc(36,0){7.5}{0.8}
\Text(36,0)[c]{$\scriptstyle{\Delta}$}
\Text(36,-14)[c]{$\scriptstyle{[n_2]}$}
\Vertex(28,0){1.8}
\ArrowLine(85,25)(60,0)
\ArrowLine(60,0)(85,-25)

\Text(100,0)[l]{, $\dots$}

\epi
&\qquad&\qquad\qquad\qquad
\nonumber \\
\nonumber \\
\nonumber \\
\big[{\cal R}^{[n]}_{1\nu}\big]_{s, {\rm I}}^{{\rm ext}}\subset\quad
\bpi(0,0)(0,-3)

\Photon(-5,25)(20,0){1.5}{5}
\Text(0,26)[l]{$\scriptstyle{\nu}$}
\GCirc(5.5,14.5){5}{0.8}
\Text(5.5,14.5)[c]{$\scriptstyle{\Delta}$}
\Photon(-5,-25)(20,0){1.5}{5}
\ArrowLine(53.5,25)(28.5,0)
\ArrowLine(28.5,0)(53.5,-25)
\GCirc(20,0){7.5}{0.8}
\Text(20,0)[c]{\scriptsize{1PI}}
\Vertex(28,0){1.8}
\Text(20,14)[c]{$\scriptstyle{[n]}$}

\Text(70,0)[l]{, $\dots$}

\epi
&\qquad&\qquad\qquad\qquad
\nonumber \\
\nonumber \\
\nonumber \\
\big[{\cal R}^{[n]}_{1\nu}\big]_{t, {\rm I}}^{{\rm int}}\subset\quad
\bpi(0,0)(0,-3)

\Photon(-5,25)(20,0){1.5}{5}
\Text(0,26)[l]{$\scriptstyle{\nu}$}
\GCirc(5.5,14.5){5}{0.8}
\Text(5.5,14.5)[c]{$\scriptstyle{\Delta}$}
\Photon(-5,-25)(20,0){1.5}{5}
\GCirc(20,0){7.5}{0.8}
\Text(20,0)[c]{\scriptsize{1PI}}
\Text(20,14)[c]{$\scriptstyle{[n_1]}$}
\ArrowLine(61,25)(36,0)
\ArrowLine(36,0)(61,-25)
\GCirc(36,0){7.5}{0.8}
\Text(36,0)[c]{\scriptsize{1PI}}
\Text(36,-14)[c]{$\scriptstyle{[n_3]}$}
\Vertex(28,0){1.8}

\Text(70,0)[l]{, $\dots$}

\epi
&\qquad&\qquad\qquad\qquad \\
\nonumber 
\eea
where the black dot indicates that a tree-level propagator has been
removed through pinching. The structure of the $[{\cal
R}^{ab}_{1\nu}]_{s,{\rm Y}}$ terms is very characteristic:
kinematically they only depend on $s$; whereas this is obviously true
for the first two equations in (\ref{xYres})
 (since it comes from the action of 
$k_1^\mu$ on an $s$-dependent piece $[{\cal
T}^{ab}_{\mu\nu}]_{s,{\rm Y}}$), is 
far less obvious for the third and fourth equations, 
since it stems from the action of 
$k_1^\mu$ on an $t$-dependent piece $[{\cal
T}^{ab}_{\mu\nu}]_{t,{\rm Y}}$. 
In addition, all the ${\cal R}$ pieces share a common feature, {\it
i.e.}, that they  
{\it cannot} be written in terms of conventional
Feynman rules; instead they are built from unphysical vertices 
which do not correspond to any term in the original
Lagrangian. All these terms are precisely the ones that cancel {\it
diagrammatically} against each other when carrying out the PT procedure.

Thus, after the PT cancellations have been enforced, we find that 
for the 1PI $t$-channel part of the amplitude, we have the equality 
\bea
[k^\mu_1 {\cal T}^{ab}_{\mu\nu}]
^{\scriptscriptstyle{\rm PT}}_{t,\scriptscriptstyle{\rm I}} &=&  
[{\cal S}_{1\nu}^{ab}]_{t,\scriptscriptstyle{\rm
I}},  \nonumber \\
\big[k^\nu_2{\cal T}^{ab}_{\mu\nu}\big]
^{\scriptscriptstyle{\rm PT}}_{t,\scriptscriptstyle{\rm I}} &=&  
[{\cal S}_{2\mu}^{ab}]_{t,\scriptscriptstyle{\rm
I}}.
\label{eq}
\eea
What is crucial in the above result is that it automatically
takes care of both the  $s$-$t$ as well as the 1PR and 1PI
cancellations of the ${\cal R}$ terms, which is  characteristic
of the PT, without having to actually trace them down. 
Thus, on hindsight, 
the PT algorithm as applied in the past
has amounted 
to enforcing diagrammatically (essentially by hand) the 
vast, BRST-driven $s$-$t$ channel cancellations, without    
making use of the all-order STIs.   
Evidently, tracing down the action of the longitudinal diagrams 
and the resulting exchanges  
between vertex and self-energy graphs, 
is equivalent to deriving (loop-by-loop) Eq.(\ref{eq}). 
Therefore,
the non-trivial step for generalizing the PT to all orders is to 
recognize that the result obtained after 
the implementation of the PT algorithm 
on the  left  hand-side of  Eq.(\ref{eq}) is already encoded  
on the right-hand  side!
Indeed, the
right-hand side involves  only conventional (ghost) Green's functions,
expressed  in terms  of normal  Feynman  rules, with  no reference  to
unphysical  vertices.   

This last point merits some additional comments.  
It should be clear that no pinching is possible when looking at the
$t$-channel irreducible part of the right  hand-side of Eq.(\ref{onshSTI}). 
So, if we were to enforce  
the  PT cancellations on both sides of the $t$ irreducible part of 
Eq.(\ref{onshSTI}), on the 
right  hand-side there would be nothing to pinch (all the vertices are
internal), and therefore,
there would be no unphysical vertices generated. Therefore,
on the left-hand side, where pinching is possible, 
all contributions containing unphysical vertices must cancel.
The only way to distort this equality is to violate the PT rules,
allowing for example the generation of additional 
longitudinal momenta by carrying out sub-integrations, or 
by splitting internal vertices. Violating 
these rules will result in undesirable consequences: 
in the first case the appearance of terms of the form 
$q\cdot k$ in the denominators will interfere with the 
manifest analyticity of 
the PT Green's functions constructed, whereas, in the second, the 
special unitarity properties of the  PT Green's functions will 
be inevitably compromised. 
 
In the next section we will study in detail how to use  
the considerations presented here in order to accomplish  
the all-order PT construction.

\section{\label{secc:allord} The PT gluon--quark--anti-quark 
vertex to all orders}

This section contains one of the central result 
of the present paper, namely the all-order
PT construction of the gluon--quark--anti-quark vertex, 
with the gluon off-shell and the quarks ``on-shell''. 
By virtue of the observations  made in the previous section, 
the derivation presented here turns 
out to be particularly compact. 
 
Before entering into the actual construction, some 
additional comments are in order. 
Once  the effective Green's functions have  been derived, they
will be  compared to the  corresponding Green's functions  obtained in
the Feynman  gauge of  the background field method   
\cite{Dewitt:ub}. 
The  latter is  a  special gauge-fixing 
procedure, implemented  at the  level of  the generating functional.  In 
particular, it preserves  the symmetry of  the action under ordinary  gauge
transformations  with respect to  the background (classical) gauge field
$\widehat{A}^a_{\mu}$,  while the quantum gauge fields $A^a_{\mu}$ appearing in
the  loops transform  homogeneously under  the gauge  group,  {\it i.e.},
as ordinary matter  fields which happened  
to be assigned to  the adjoint representation 
\cite{Weinberg:kr}.   As a  result  of the  background gauge symmetry, the 
$n$-point  functions  $\langle 0  | T \left[ \widehat{A}^{a_1}_{\mu_1}(x_1)
\widehat{A}^{a_2}_{\mu_2}(x_2)\dots  
\widehat{A}^{a_n}_{\mu_n}(x_n) \right] |0 \rangle$
are gauge-invariant, in the sense that they 
satisfy naive, QED-like WIs.
Notice however that they are {\it not} gauge-independent, 
because they 
depend {\it explicitly} on the quantum gauge-fixing parameter 
$\xi_Q$ used  to define the tree-level  propagators of  the 
quantum gluons.
In theories with 
spontaneous symmetry breaking this dependence on $\xi_Q$ gives rise 
to {\it unphysical} thresholds inside these Green's functions 
for $\xi_Q \neq 1$, a fact
which limits their usefulness for resummation purposes 
\cite{Papavassiliou:1995fq}.
Only the case of the background Feynman gauge (BFG)
({\it i.e.} background field method with $\xi_Q =1$) 
is free from unphysical poles, 
and the results of these Green's functions collapse to those of the PT, 
at one- \cite{Denner:1994nn}
and two-loops \cite{Papavassiliou:2000az}.
As we will see, this 
correspondence between the PT Green's functions and the ones
obtained using the BFG persists  to all
orders. This fact provides a valuable book-keeping scheme, since, 
once the  
equality between the Green's functions obtained 
using either schemes has been established 
(and {\it only} then), the background field method 
Feynman rules may be directly 
employed. 
We note in passing that  the PT construction  goes through
unaltered under  circumstances  where  the background  field  method
Feynman rules cannot even be  applied.  Specifically, if instead of an
$S$-matrix element  one were to consider a  different observable, such
as a current-current correlation function  or a Wilson loop (as was in
fact    done    by    Cornwall    in    the    original    formulation
\cite{Cornwall:1982zr}, and more recently in \cite{Binosi:2001hy}) one
could not start  out using the background Feynman  rules, because {\it
all} fields  appearing inside the  first non-trivial loop  are quantum
ones. Instead, by following the PT rearrangement inside these physical
amplitudes one would ``dynamically'' arrive at the BFG answer.
After these clarifying comments 
we proceed with the actual all-order construction.

To  begin with,  it is  immediate  to recognize  that, in  the RFG,  
box diagrams of arbitrary order $n$,  to be denoted by $B^{[n]}$, coincide
with the PT boxes ${\widehat B}^{[n]}$, since all three-gluon vertices
are  ``internal'',  {\it  i.e.},   they  do  not  provide  longitudinal
momenta. Thus, they coincide  with the BFG boxes, $\tilde{B}^{[n]}$,
{\it  i.e.},
\be
{\widehat B}^{[n]}  = B^{[n]}  =  \tilde{B}^{[n]}
\label{res2:boxes}
\ee
for every~$n$. The same is true for the PT quark self-energies;  
for exactly the same reason, they coincide with their RFG (and BFG)  
counterparts, {\it  i.e.}  
\be
{\widehat \Sigma}^{ab \,[n]}  = \Sigma^{ab \,[n]}  =  
\tilde{\Sigma}^{ab \,[n]} . 
\label{res2:sigmas}
\ee

The next step will be  the construction of the all orders PT
gluon--quark--anti-quark 1PI vertex 
$\widehat\Gamma^{e}_{\alpha}(q,\pci)$. We start by  
classifying all the diagrams that contribute to this
vertex in the RFG, into the following categories: ({\it i}) 
those containing an
external three-gluon vertex, {\it i.e.}, those containing a
three-gluon vertex where the momentum $q$ is incoming, and ({\it ii})
those which do not have such an external three-gluon vertex. This
latter set contains graphs where the incoming gluon couples to the
rest of the diagram with any other type of interaction vertex other
than a three-gluon vertex. Thus we write \cite{Papavassiliou:2000az}  
\be
\Gamma^{e}_{\alpha}(q,\pci)=
\Gamma^{e}_{\!A^3,\alpha}(q,\pci)+
\Gamma^{e}_{\!A^4,\alpha}(q,\pci)+
\Gamma^{e}_{\!A\bar c c,\alpha}(q,\pci)+
\Gamma^{e}_{\!A\bar q q,\alpha}(q,\pci).
\ee
Then, the above Green's functions have the following diagrammatic
representation
%
%A2 and A3
%
\bce
\bpi(0,50)(130,-20)

\Text(-25,-7.5)[r]{$\Gamma^e_{\!A^3,\alpha}(q,\pci)=$}

\Photon(-15,-7)(1,-7){1.5}{3}
\Text(-15,-13)[l]{$\scriptstyle{\alpha}$}
\Text(-15,-1)[l]{$\scriptstyle{e}$}

\PhotonArc(17.5,-7.5)(15,0,360){1.5}{17}
\GCirc(17.5,6.5){5}{0.8}
\Text(17.5,6.5)[c]{$\scriptstyle{\Delta}$}
\Text(5,7.5)[c]{$\scriptstyle{\nu}$}
\Text(32.5,7.5)[c]{$\scriptstyle{\sigma}$}
\GCirc(17.5,-22.5){5}{0.8}
\Text(17.5,-22.5)[c]{$\scriptstyle{\Delta}$}
\Text(5,-22.5)[c]{$\scriptstyle{\mu}$}
\Text(32.5,-22.5)[c]{$\scriptstyle{\rho}$}

%\ArrowLine(60,10)(40,-5)
%\ArrowLine(75,20)(60,10)
%\GCirc(58.5,8.5){5}{0.8}
%\Text(58.5,8.5)[c]{$\scriptstyle{S}$}
\ArrowLine(65,20)(38,-7)
\Text(70,20)[l]{$\scriptstyle{p_1}$}

%\ArrowLine(40,-10)(60,-25)
%\ArrowLine(60,-25)(75,-35)
%\GCirc(58.5,-23.5){5}{0.8}
%\Text(58.5,-23.5)[c]{$\scriptstyle{S}$}
\ArrowLine(38,-7)(65,-34)
\Text(70,-34)[l]{$\scriptstyle{p_2}$}

\GCirc(37.5,-7.5){10}{0.8}
\Text(37,-7.5)[c]{${\cal C}^{\scriptscriptstyle{{t,{\rm
I}}}}_{\rho\sigma}$} 

\Text(215,-7.5)[r]{$\Gamma^e_{\!A^4,\alpha}(q,\pci)=$}

\PhotonArc(257.5,-7.5)(15,0,360){1.5}{17}
\GCirc(257.5,6.5){5}{0.8}
\Text(258.5,6.5)[c]{$\scriptstyle{\Delta}$}
\Text(245,7.5)[c]{$\scriptstyle{\nu}$}
\Text(272.5,7.5)[c]{$\scriptstyle{\sigma}$}
\GCirc(257.5,-22.5){5}{0.8}
\Text(258.5,-22.5)[c]{$\scriptstyle{\Delta}$}
\Text(245,-22.5)[c]{$\scriptstyle{\mu}$}
\Text(272.5,-22.5)[c]{$\scriptstyle{\rho}$}

\Photon(225,-7)(270,-7){-1.5}{8}
\GCirc(256.5,-7){5}{0.8}
\Text(257.5,-7)[c]{$\scriptstyle{\Delta}$}
\Text(225,-13)[l]{$\scriptstyle{\alpha}$}
\Text(225,-1)[l]{$\scriptstyle{e}$}
\Text(250,-12)[c]{$\scriptscriptstyle{\chi}$}
\Text(265.5,-1)[c]{$\scriptstyle{\lambda}$}

%\ArrowLine(300,10)(280,-5)
%\ArrowLine(315,20)(300,10)
%\GCirc(298.5,8.5){5}{0.8}
%\Text(298.5,8.5)[c]{$\scriptstyle{S}$}
\ArrowLine(305,20)(278,-7)
\Text(310,20)[l]{$\scriptstyle{p_1}$}

%\ArrowLine(280,-10)(300,-25)
%\ArrowLine(300,-25)(315,-35)
%\GCirc(298.5,-23.5){5}{0.8}
%\Text(298.5,-23.5)[c]{$\scriptstyle{S}$}
\ArrowLine(278,-7)(305,-34)
\Text(310,-34)[l]{$\scriptstyle{p_2}$}

\GCirc(277.5,-7.5){10}{0.8}
\Text(279,-7.5)[c]{${{\cal C}^{\scriptscriptstyle{{t,{\rm
I}}}}_{\!\!\rho\lambda\sigma}}$}
\epi
\ece
%
%cbar c
%
\bce
\bpi(0,70)(85,-20)

\Text(-25,-7.5)[r]{$\Gamma^e_{\!A\bar c c,\alpha}(q,\pci)=$}

\Photon(-15,-7)(2.5,-7){1.5}{3}
\Text(-15,-13)[l]{$\scriptstyle{\alpha}$}
\Text(-15,-1)[l]{$\scriptstyle{e}$}

\DashCArc(17.5,-7.5)(15,0,360){1}
\GCirc(17.5,6.5){5}{0.8}
\Text(17.5,6.5)[c]{$\scriptstyle{D}$}
\DashArrowLine(7,3.2)(6.9,3.1){1}
\DashArrowLine(6.9,-18.1)(7,-18.2){1}
\DashArrowLine(28.1,-18.2)(28.2,-18.1){1}
\DashArrowLine(28.2,3.1)(28.1,3.2){1}
\GCirc(17.5,-22.5){5}{0.8}
\Text(17.5,-22.5)[c]{$\scriptstyle{D}$}

%\ArrowLine(60,10)(40,-5)
%\ArrowLine(75,20)(60,10)
%\GCirc(58.5,8.5){5}{0.8}
%\Text(58.5,8.5)[c]{$\scriptstyle{S}$}
\ArrowLine(65,20)(38,-7)
\Text(70,20)[l]{$\scriptstyle{p_1}$}

%\ArrowLine(40,-10)(60,-25)
%\ArrowLine(60,-25)(75,-35)
%\GCirc(58.5,-23.5){5}{0.8}
%\Text(58.5,-23.5)[c]{$\scriptstyle{S}$}
\ArrowLine(38,-7)(65,-34)
\Text(70,-34)[l]{$\scriptstyle{p_2}$}

\GCirc(37.5,-7.5){10}{0.8}
\Text(37,-7.5)[c]{${\cal G}^{\scriptscriptstyle{t,{\rm
I}}}_{\scriptscriptstyle{1}}$}

\Text(110,-7)[c]{$+$}

\Photon(135,-7)(152.5,-7){1.5}{3}
\Text(135,-13)[l]{$\scriptstyle{\alpha}$}
\Text(135,-1)[l]{$\scriptstyle{e}$}

\DashCArc(167.5,-7.5)(15,0,360){1}
\GCirc(167.5,6.5){5}{0.8}
\Text(167.5,6.5)[c]{$\scriptstyle{D}$}
\DashArrowLine(156.9,3.1)(157,3.2){1}
\DashArrowLine(157,-18.2)(156.9,-18.1){1}
\DashArrowLine(178.2,-18.1)(178.1,-18.2){1}
\DashArrowLine(178.1,3.2)(178.2,3.1){1}
\GCirc(167.5,-22.5){5}{0.8}
\Text(167.5,-22.5)[c]{$\scriptstyle{D}$}

%\ArrowLine(210,10)(190,-5)
%\ArrowLine(225,20)(210,10)
%\GCirc(208.5,8.5){5}{0.8}
%\Text(208.5,8.5)[c]{$\scriptstyle{S}$}
\ArrowLine(215,20)(188,-7)
\Text(220,20)[l]{$\scriptstyle{p_1}$}

%\ArrowLine(190,-10)(210,-25)
%\ArrowLine(210,-25)(225,-35)
%\GCirc(208.5,-23.5){5}{0.8}
%\Text(208.5,-23.5)[c]{$\scriptstyle{S}$}
\ArrowLine(188,-7)(215,-34)
\Text(220,-34)[l]{$\scriptstyle{p_2}$}

\GCirc(187.5,-7.5){10}{0.8}
\Text(187.5,-7.5)[c]{${\cal G}^{\scriptscriptstyle{t,{\rm
I}}}_{\scriptscriptstyle{2}}$}
%
%qbar q
%
\epi
\ece

\bce
\bpi(0,85)(85,-35)

\Text(-25,-7.5)[r]{$\Gamma^e_{\!A\bar q q,\alpha}(q,\pci)=$}

\Photon(-15,-7)(2.5,-7){1.5}{3}
\Text(-15,-13)[l]{$\scriptstyle{\alpha}$}
\Text(-15,-1)[l]{$\scriptstyle{e}$}

\CArc(17.5,-7.5)(15,0,360)
\GCirc(17.5,6.5){5}{0.8}
\Text(17.5,6.5)[c]{$\scriptstyle{S}$}
\DashArrowLine(7,3.2)(6.9,3.1){1}
\DashArrowLine(6.9,-18.1)(7,-18.2){1}
\DashArrowLine(28.1,-18.2)(28.2,-18.1){1}
\DashArrowLine(28.2,3.1)(28.1,3.2){1}
\GCirc(17.5,-22.5){5}{0.8}
\Text(17.5,-22.5)[c]{$\scriptstyle{S}$}

%\ArrowLine(60,10)(40,-5)
%\ArrowLine(75,20)(60,10)
%\GCirc(58.5,8.5){5}{0.8}
%\Text(58.5,8.5)[c]{$\scriptstyle{S}$}
\ArrowLine(65,20)(38,-7)
\Text(70,20)[l]{$\scriptstyle{p_1}$}

%\ArrowLine(40,-10)(60,-25)
%\ArrowLine(60,-25)(75,-35)
%\GCirc(58.5,-23.5){5}{0.8}
%\Text(58.5,-23.5)[c]{$\scriptstyle{S}$}
\ArrowLine(38,-7)(65,-34)
\Text(70,-34)[l]{$\scriptstyle{p_2}$}

\GCirc(37.5,-7.5){10}{0.8}
\Text(36.5,-7.5)[c]{${\cal H}^{\scriptscriptstyle{t,{\rm
I}}}_{\scriptscriptstyle{1}}$}

\Text(110,-7)[c]{$+$}

\Photon(135,-7)(152.5,-7){1.5}{3}
\Text(135,-13)[l]{$\scriptstyle{\alpha}$}
\Text(135,-1)[l]{$\scriptstyle{e}$}

\CArc(167.5,-7.5)(15,0,360)
\GCirc(167.5,6.5){5}{0.8}
\Text(167.5,6.5)[c]{$\scriptstyle{S}$}
\DashArrowLine(156.9,3.1)(157,3.2){1}
\DashArrowLine(157,-18.2)(156.9,-18.1){1}
\DashArrowLine(178.2,-18.1)(178.1,-18.2){1}
\DashArrowLine(178.1,3.2)(178.2,3.1){1}
\GCirc(167.5,-22.5){5}{0.8}
\Text(167.5,-22.5)[c]{$\scriptstyle{S}$}

%\ArrowLine(210,10)(190,-5)
%\ArrowLine(225,20)(210,10)
%\GCirc(208.5,8.5){5}{0.8}
%\Text(208.5,8.5)[c]{$\scriptstyle{S}$}
\ArrowLine(215,20)(188,-7)
\Text(220,20)[l]{$\scriptstyle{p_1}$}

%\ArrowLine(190,-10)(210,-25)
%\ArrowLine(210,-25)(225,-35)
%\GCirc(208.5,-23.5){5}{0.8}
%\Text(208.5,-23.5)[c]{$\scriptstyle{S}$}
\ArrowLine(188,-7)(215,-34)
\Text(220,-34)[l]{$\scriptstyle{p_2}$}

\GCirc(187.5,-7.5){10}{0.8}
\Text(187,-7.5)[c]{${\cal H}^{\scriptscriptstyle{t,{\rm
I}}}_{\scriptscriptstyle{2}}$} 

\epi
\ece

As a second step, we next carry out inside the class ({\it i})
diagrams 
the vertex decomposition given in Eq.(\ref{decomp}); thus we write 
\be
\Gamma^{e}_{\!A^3,\alpha}(q,\pci)=
\Gamma^{{\rm F\!,\,}e}_{\!A^3,\alpha}(q,\pci)+
\Gamma^{{\rm P\!,\,}e}_{\!A^3,\alpha}(q,\pci),
\ee
where
\bea
\Gamma^{{\rm F\!,\,}e}_{\!A^3,\alpha}(q,\pci)&=&
gf^{eba}\int\!\Gamma^{{\rm F\!,\,}\nu\mu}_\alpha(q,-k,k-q)
\left[{\cal T}_{\mu\nu}^{ab}(-k+q,k,\pci)
\right]_{t,{\rm I}}, \nonumber \\
\Gamma^{{\rm P\!,\,}e}_{\!A^3,\alpha}(q,\pci)&=&
gf^{eba}\int\!\left[(k-q)^\mu g^\nu_\alpha+k^\nu g^\mu_\alpha\right]
\left[{\cal T}_{\mu\nu}^{ab}(-k+q,k,\pci)
\right]_{t,{\rm I}},
\eea
and we have defined the integral measure
\be
\int\equiv\mu^{2\varepsilon}\int\!\frac{d^dk}{(2\pi)^d},
\label{im}
\ee
with $d=4-2\varepsilon$ and $\mu$ the 't~Hooft mass.
Following the discussion presented in the previous section,
the  pinching   action  amounts  to  the   replacement  
\be
k^\nu  [{\cal
T}_{\mu\nu}^{ab}]_{t,{\rm I}}(-k+q,k,\pci) \to[k^\nu{\cal
T}_{\mu\nu}^{ab}(-k+q,k,\pci)]_{t,{\rm  I}}  = \left[{\cal
S}_{2\mu}^{ab}(-k+q,k,\pci)\right]_{t,{\rm  I}}
\ee
and  similarly  for the
term coming from the momentum 
$(k-q)^\mu$,  or,    equivalently, 
\be    
\Gamma^{{\rm
P\!,\,}e}_{\!A^3,\alpha}(q,\pci) \rightarrow  gf^{eba} \int\!
\left\{\left[{\cal
S}_{2\alpha}^{ab}(-k+q,k,\pci)\right]_{t,{\rm  I}}  - \left[{\cal
S}_{1\alpha}^{ab}(-k+q,k,\pci)\right]_{t,{\rm I}}\right\}.
\label{PTvertex}
\ee

At   this  point  the   construction  of   the  effective   PT  vertex
$\widehat\Gamma^{e}_{\alpha}$ has been  completed, and we have 
\bea
\widehat\Gamma^{e}_{\alpha}(q,\pci)&=&
\Gamma^{{\rm F\!,}\,e}_{\!A^3,\alpha}(q,\pci)+
\Gamma^{e}_{\!A^4,\alpha}(q,\pci)+
\Gamma^{e}_{\!A\bar c c,\alpha}(q,\pci)+
\Gamma^{e}_{\!A\bar q q,\alpha}(q,\pci) \nonumber \\
&+&gf^{eba}\int\!\left\{
\left[{\cal S}_{2\alpha}^{ab}(-k+q,k,\pci)\right]_{t,{\rm
I}}-\left[{\cal S}_{1\alpha}^{ab}(-k+q,k,\pci)\right]_{t,{\rm
I}}\right\}.
\eea
We emphasize that 
in the construction presented thus far we have
never resorted to the background  
formalism, but have only used the BRST identities of Eq.(\ref{onshSTI}),
together with Eq.(\ref{eq}). 
The next crucial question will be then to establish the
connection between the effective PT vertex and the
gluon--quark--anti-quark vertex 
$\widetilde\Gamma^{e}_{\alpha}(q,\pci)$ written in the BFG.
For doing this we first of all observe that all the ``inert'' terms
contained in the original RFG $\Gamma^{e}_{\alpha}(q,\pci)$ vertex carry
over the same sub-groups of BFG graphs.
In order to facilitate this identification  
we remind the reader that in the background field method 
the bare three- and four-gluon vertices involving background 
and quantum gluons depend on the gauge-fixing parameter 
$\xi_Q$. In particular, the former involving one
background gluon (carrying four-momentum
$q$) and two quantum ones (carrying four-momenta $k_1$ and $k_2$),
reads
\be
\widetilde\Gamma_{\alpha\mu\nu}(q,k_1,k_2)=(q-k_1-\frac1{\xi_Q}k_2)_\nu
g_{\alpha\mu}+(k_1-k_2)_{\alpha}g_{\mu\nu}+(k_2-q+\frac1{\xi_Q}k_1)_\mu
g_{\alpha\nu},
\label{BFMbare}
\ee
which can be rewritten as
\be
\widetilde\Gamma_{\alpha\mu\nu}(q,k_1,k_2)=
\Gamma^{\rm
  F}_{\alpha\mu\nu}(q,k_1,k_2)-\left(\frac{1-\xi_Q}{\xi_Q}\right)
\Gamma^{\rm P}_{\alpha\mu\nu}(q,k_1,k_2),
\label{firstdec}
\ee
or
\be
\widetilde\Gamma_{\alpha\mu\nu}(q,k_1,k_2)=
\Gamma_{\alpha\mu\nu}(q,k_1,k_2)-\frac1{\xi_Q}
\Gamma^{\rm P}_{\alpha\mu\nu}(q,k_1,k_2).
\label{secondec}
\ee
Eq.(\ref{firstdec}) implies then that in the BFG ($\xi_Q =1$)
the bare vertex of Eq.(\ref{BFMbare})  
coincides with the $\Gamma_{\alpha \mu \nu}^{{\rm F}}(q,k_1,k_2)$  
of Eq.(\ref{GFGP}).
Similarly, 
the four-particle vertex involving two background and two quantum
gluons reduces at $\xi_Q =1$  to the usual four-gluon vertex. 
Thus we have
\bea
\Gamma^{{\rm F\!,}\,e}_{\!A^3,\alpha}(q,\pci)&\equiv&
\widetilde\Gamma^{e}_{\!\widetilde AA^2,\alpha}(q,\pci), \nonumber \\
\Gamma^{e}_{\!A^4,\alpha}(q,\pci)&\equiv&
\widetilde\Gamma^{e}_{\!\widetilde AA^3,\alpha}(q,\pci), \nonumber \\
\Gamma^{e}_{\!A\bar q q,\alpha}(q,\pci)&\equiv&
\widetilde\Gamma^{e}_{\!\widetilde A\bar q q,\alpha}(q,\pci),
\eea
where $\widetilde A$ is the background gluon. 
The only exception are
the ghost diagrams $\Gamma^{e}_{\!A\bar cc,\alpha}(q,\pci)$. 
The  important step is  to recognize  that the  BFG ghost  sector is
provided  precisely  by combining  the  RFG  ghosts  with the
right-hand side  of 
Eq.(\ref{eq}).  Specifically,  one arrives  at both  the {\it  symmetric}
vertex $\widetilde\Gamma_{\widetilde{A}\bar  c c}^{e}(q,\pci)$, characteristic
of       the      BFG,      as       well      as       at    the
four-particle ghost vertex
$\widetilde\Gamma_{\!\widetilde{A} A\bar  c c}^{e}(q,\pci)$, with
\bce
\bpi(0,50)(110,-30)

\Text(15,-7)[r]{$\widetilde\Gamma^e_{\!\widetilde A A\bar cc,\alpha}(q,\pci)=$}

\PhotonArc(57.5,-7.5)(15,0,180){-1.5}{8.5}
\GCirc(57.5,6.5){5}{0.8}
\Text(57.5,6.5)[c]{$\scriptstyle{\Delta}$}
\Text(45,7.5)[c]{$\scriptstyle{\nu}$}
\Text(72.5,7.5)[c]{$\scriptstyle{\sigma}$}
\DashCArc(57.5,-7.5)(16.5,180,0){1}
\GCirc(57.5,-22.5){5}{0.8}
\Text(57.5,-22.5)[c]{$\scriptstyle{D}$}
\DashArrowLine(69.2,-19.2)(69.3,-19.1){1}
\DashArrowLine(45.8,-19.1)(45.9,-19.2){1}

\DashArrowLine(50.5,-7)(41,-7){1}
\DashArrowLine(67,-7)(60.5,-7){1}
\Photon(25,-7)(41,-7){1.5}{3}
\GCirc(55.5,-7){5}{0.8}
\Text(55.5,-7)[c]{$\scriptstyle{D}$}
\Text(25,-13)[l]{$\scriptstyle{\alpha}$}
\Text(25,-1)[l]{$\scriptstyle{e}$}

\ArrowLine(105,20)(78,-7)
\Text(110,20)[l]{$\scriptstyle{p_1}$}

\ArrowLine(78,-7)(105,-34)
\Text(110,-34)[l]{$\scriptstyle{p_2}$}

\GCirc(77.5,-7.5){10}{0.8}
\Text(76.5,-7.5)[c]{${{\cal F}^{\scriptscriptstyle{{t,{\rm
I}}}}_{\!1\sigma}}$}

\Text(150,-7)[c]{$+$}

\PhotonArc(207.5,-7.5)(15,180,0){-1.5}{8.5}
\GCirc(207.5,-22.5){5}{0.8}
\Text(208,-22.5)[c]{$\scriptstyle{\Delta}$}
\Text(195,-22.5)[c]{$\scriptstyle{\mu}$}
\Text(222.5,-22.5)[c]{$\scriptstyle{\rho}$}
\DashCArc(207.5,-7.5)(16.5,0,180){1}
\GCirc(207.5,6.5){5}{0.8}
\Text(208,6.5)[c]{$\scriptstyle{D}$}
\DashArrowLine(219.2,4.2)(219.3,4.1){1}
\DashArrowLine(195.8,4.1)(195.9,4.2){1}

\DashArrowLine(200.5,-7)(191,-7){1}
\DashArrowLine(217,-7)(210.5,-7){1}
\Photon(175,-7)(191,-7){1.5}{3}
\GCirc(205.5,-7){5}{0.8}
\Text(205.5,-7)[c]{$\scriptstyle{D}$}
\Text(175,-13)[l]{$\scriptstyle{\alpha}$}
\Text(175,-1)[l]{$\scriptstyle{e}$}

\ArrowLine(255,20)(228,-7)
\Text(260,20)[l]{$\scriptstyle{p_1}$}

\ArrowLine(228,-7)(255,-34)
\Text(260,-34)[l]{$\scriptstyle{p_2}$}

\GCirc(227.5,-7.5){10}{0.8}
\Text(226.5,-7.5)[c]{${{\cal F}^{\scriptscriptstyle{{t,{\rm
I}}}}_{\!2\rho}}$}

\epi
\ece
which is totally
absent in the conventional formalism.
Indeed, using Eq.(\ref{onshdef}), we find (omitting the spinors) 
\bea
\int\!
\left[{\cal S}_{1\alpha}^{ab}(-k+q,k,\pci)\right]_{t,{\rm
I}}&=&-\int\!k_\alpha\left[
{\cal G}_1^{ab}(-k+q,k,\pci)\right]_{t,{\rm
I}}D(-k+q)D(k)\nonumber \\
&+& igf^{bcd}\!\int\!
\left[
{\cal Q}_{1\alpha}^{acd}(-k+q,k,\pci)\right]_{t,{\rm
I}}D(-k+q),\nonumber \\
\int\!
\left[{\cal S}_{2\alpha}^{ab}(-k+q,k,\pci)\right]_{t,{\rm
I}}&=&\int\!(k-q)_\alpha\left[
{\cal G}_2^{ab}(-k+q,k,\pci)\right]_{t,{\rm
I}}D(-k+q)D(k)\nonumber \\
&+&igf^{acd\!}\int\!
\left[
{\cal Q}_{2\alpha}^{cdb}(-k+q,k,\pci)\right]_{t,{\rm
I}}D(k).
\eea
Then it is easy to check that
\bea
\widetilde\Gamma^{e}_{\!\widetilde A\bar c
c,\alpha}(q,\pci)&=&
\Gamma^{e}_{\!A\bar c
c,\alpha}(q,\pci)+gf^{eba}\int\!k_\alpha\left[ 
{\cal G}_1^{ab}(-k+q,k,\pci)\right]_{t,{\rm
I}}D(-k+q)D(k) \nonumber \\
&+& gf^{eba}\int\!(k-q)_\alpha\left[
{\cal G}_2^{ab}(-k+q,k,\pci)\right]_{t,{\rm
I}}D(-k+q)D(k), \nonumber \\
\widetilde\Gamma^{e}_{\!\widetilde AA\bar c
c,\alpha}(q,\pci)&=&-ig^2f^{eba}f^{bcd}\int\!
\left[
{\cal Q}_{1\alpha}^{acd}(-k+q,k,\pci)\right]_{t,{\rm
I}}D(-k+q) \nonumber \\
&+&ig^2f^{eba}f^{acd}\int\!
\left[
{\cal Q}_{2\alpha}^{cdb}(-k+q,k,\pci)\right]_{t,{\rm
I}}D(k),
\eea
which gives us the final identity
\be
\widehat\Gamma^{e}_{\alpha}(q,\pci) \equiv
\widetilde\Gamma^{e}_{\alpha}(q,\pci).
\label{res1:vertex}
\ee

Once again, we emphasize that
the  sole ingredient in  the above  construction has  been the  STIs of
Eq.(\ref{onshSTI});  in particular,  at  no point  have  we employed  {\it
a priori} the  background field method formalism.  
Instead, its  special  ghost sector
has  arisen {\it  dynamically},  once the  PT  rearrangement has  taken
place.
An immediate consequence of the above correspondence between PT and BFG is 
that  $\widehat\Gamma^{e}_{\alpha}(q,\pci)$ satisfies the QED-like WI 
\be
q^{\alpha} \widehat\Gamma^{e}_{\alpha}(q,p_1,p_2) = 
f^{ebc}\bigg[\widehat{\Sigma}^{bc} (p_1) - \widehat{\Sigma}^{bc} (p_2)\bigg].
\label{GS}
\ee

The final  step, to be undertaken in detail in the next section,  
is  to construct the  all orders PT  gluon self-energy
$\widehat\Pi_{\mu\nu}^{ab}(q)$.  Notice  that at this  point one would
expect  that  it  too  coincides  with the  BFG  gluon
self-energy  $\widetilde\Pi^{ab}_{\mu\nu}(q)$,  since  both the  boxes
${\widehat  B}$  and  the vertex  $\widehat\Gamma^{e}_{\alpha}(q,\pci)$  do
coincide with  the corresponding BFG boxes ${\widetilde
B}$ and vertex  $\widetilde\Gamma^{e}_{\alpha}(q,\pci)$, and the $S$-matrix
is unique. We will end this section showing  that
this  is indeed the case. 
To that end we will present a more detailed version 
of a proof based
on  the strong induction principle, which 
first appeared in \cite{Binosi:2002ft}. This principle   
states that a given 
predicate $P(n)$ on $\mathbb N$ is true $\forall\ n\in{\mathbb N}$, 
if $P(k)$ is true whenever $P(j)$ is true $\forall\  j\in{\mathbb N}$
with $j<k$.  

In order to avoid notational clutter, we will use the schematic notation
introduced  in  Eq.(\ref{schemnot}), suppressing all the group, Lorentz, 
and momentum indices. At  one-  and  two-loop ({\it  i.e.},
$n=1,2$), we know that the result is true due to explicit calculations
\cite{Cornwall:1982zr,Papavassiliou:2000az}.  Let  us then assume that
the PT construction has been  successfully carried out up to the order
$n-1$ (strong  induction hypothesis):  we will show  then that  the PT
gluon self-energy is equal to  the BFG gluon self-energy
at order~$n$, {\it i.e.}, $\widehat\Pi^{[n]}\equiv\widetilde\Pi^{[n]}$;  
hence,  by  the  strong
induction principle, this last result  will be valid at any given $n$,
showing finally that the PT construction holds true to all orders.

From the strong inductive hypothesis, we know that
\bea
\widehat\Pi^{[\ell]}&\equiv&\widetilde\Pi^{[\ell]}, \nonumber \\
\widehat\Gamma^{[\ell]}&\equiv&\widetilde\Gamma^{[\ell]}, \nonumber \\
\widehat B^{[\ell]}&\equiv&\widetilde B^{[\ell]}\equiv B^{[\ell]},
\label{hypo}
\eea
where $\ell=1,\dots,n-1$. 

Now, the $S$-matrix element of order $n$, to be denoted as $S^{[n]}$,
assumes the form 
\be
S^{[n]}=\left\{\Gamma\Delta\Gamma\right\}^{[n]}+B^{[n]}.
\ee 
Moreover, since it is unique, regardless if it is written in the Feynman
gauge, in the BFG, as well as before and after the PT
rearrangement, we have that \mbox{$S^{[n]}\equiv
\widehat S^{[n]}\equiv\widetilde 
S^{[n]}$}. Using then
Eq.(\ref{res2:boxes}) (which is all orders, implying that 
the last equation
in (\ref{hypo}) holds true also when $\ell=n$), we find that
\be
\left\{\Gamma\Delta\Gamma\right\}^{[n]}\equiv
\{\widehat\Gamma\widehat\Delta\widehat\Gamma\}^{[n]}\equiv
\{\widetilde\Gamma\widetilde\Delta\widetilde\Gamma\}^{[n]}.
\label{fun1}
\ee

The above amplitudes can then be split into 1PR and 1PI parts; in
particular, due to the strong inductive 
hypothesis of Eq.(\ref{hypo}) the 1PR part after 
the PT rearrangement coincides with the 1PR part written in the 
BFG since
\be
\left\{\Gamma\Delta\Gamma\right\}^{[n]}_{\scriptscriptstyle{\rm R}}=
\Gamma^{[n_1]}\Delta^{[n_2]}\Gamma^{[n_3]}, \qquad \left\{
\begin{array}{l}
n_1,n_2,n_3<n, \\
n_1+n_2+n_3=n. 
\end{array}
\right.
\ee

Then Eq.(\ref{fun1}) state the equivalence of the 1PI parts, {\it i.e.},
\be
\{\widehat\Gamma\widehat\Delta\widehat\Gamma\}_{\scriptscriptstyle{\rm
I}}^{[n]}\equiv 
\{\widetilde\Gamma\widetilde\Delta\widetilde\Gamma\}_{\scriptscriptstyle{\rm
I}}^{[n]}, 
\ee
which implies
\be
\left(\widehat\Gamma^{[n]}-\widetilde\Gamma^{[n]}\right)
\Delta^{[0]}\Gamma^{[0]}+\Gamma^{[0]}\Delta^{[0]}\left(\widehat\Gamma^{[n]}
-\widetilde\Gamma^{[n]}\right)+\Gamma^{[0]}\Delta^{[0]}
\left(\widehat\Pi^{[n]}-\widetilde\Pi^{[n]}\right)
\Delta^{[0]}\Gamma^{[0]}\equiv0.
\ee
At this point we do  not have the equality we want yet, but only that
\bea
\widehat\Gamma^{[n]}&=&\widetilde\Gamma^{[n]}+{\cal K}^{[n]}\Gamma^{[0]},
\nonumber \\
\widehat\Pi^{[n]}&=&\widetilde\Pi^{[n]}-2iq^2{\cal K}^{[n]},
\eea
with ${\cal K}^{[n]}$ an arbitrary function of $q^2$. However, by means of the
{\it explicit} construction of the PT vertex just presented, 
we have the all orders identity of 
Eq.(\ref{res1:vertex}),
so that the second equation in (\ref{hypo}) actually
holds true even when $\ell=n$, {\it i.e.},
$\widehat\Gamma^{[n]}\equiv\widetilde\Gamma^{[n]}$; then one
immediately gets  
\be
\widehat\Pi^{[n]}\equiv\widetilde\Pi^{[n]}.
\ee
Hence, by strong induction, the above relation is true for any given
order $n$, {\it i.e.}, inserting back the Lorentz and gauge group structures,
we arrive at the announced result
\be
\widehat\Pi^{ab}_{\mu\nu}(q)\equiv\widetilde\Pi^{ab}_{\mu\nu}(q).
\ee

In the next section we will carry out the construction of the 
PT gluon self-energy in detail, and will see how the above 
conclusion is explicitly realized. 

\section{\label{sec:IPT} The PT gluon self-energy to all orders}

As we have seen in the previous sections, and as has been explained in
detail in the literature,  when constructing the PT two-point function
various well-defined propagator-like  contributions are moved from the
three-point function to the two-point function.  These pinch terms are
always missing  one or more propagators corresponding  to the external
legs of  the two-point  function under construction.  Pictorially this
characteristic structure gives rise to the appearance of the unphysical
effective   vertices,  mentioned   earlier.   Of   course,   all  such
contributions, when re-alloted to the original two-point function will
cancel  exactly against analogous  contributions concealed  inside it.
Reversing  the order,  the  normal Feynman  diagrams  ({\it i.e.} with  both
external  legs present)  contributing to  the two-point  function must
contain  pieces  that  are  effectively proportional  to  the  inverse
propagators  of  the  external  legs,  a fact  which  allows  them  to
communicate  (and eventually  cancel) against  the pinch  parts coming
from  the three-point  function  (or  the boxes,  when  away from  the
Feynman gauge).  Thus, when constructing the PT  gluon self-energy one
may  follow  two  equivalent  procedures.  First,  one  may  determine
explicitly the pinch terms coming from  the vertex and add them to the
conventional graphs; this would correspond to the usual ``$S$-matrix''
PT  construction.  Second,  one  may  isolate  from  the  conventional
self-energy  all the  aforementioned  terms that  are proportional  to
inverse propagators, and simply discard them; this would correspond to
the ``intrinsic PT''.  In this  latter one avoids the embedding of the
PT objects  into $S$-matrix  elements, and manipulates  only off-shell
self-energy  corrections.    The  two  constructions   are  absolutely
equivalent: discarding  the aforementioned terms  in the ``intrinsic''
construction is  justified because we  know that inside  an $S$-matrix
element they  will eventually cancel  (to all orders)  against similar
pieces stemming from the vertices.  

In what follows we will present 
in detail the intrinsic construction, which, in addition to being 
more economical, it is 
intimately connected to the STI 
of $A_{\mu}^a\, A_{\nu}^b\, q^i\, \bar{q}^j$,
employed in the previous sections.  
The important point is that the characteristic terms containing 
inverse propagators  arise from the
STI satisfied by the three-gluon vertex (of arbitrary order) 
appearing inside appropriate sets of
diagrams, when it is contracted by longitudinal momenta. 
In fact, these terms are precisely the set 
of unphysical contributions 
$[{\cal R}^{ab}_{1\nu}]_{s,\scriptscriptstyle{\rm I}}^{\rm ext}$
produced by the action of a longitudinal momentum on  
the term $[{\cal T}^{ab}_{\mu\nu}]_{s,\scriptscriptstyle{\rm I}}$,
as shown in Eq.(\ref{xYres}). Evidently, the STI satisfied by 
the (full) three-gluon vertex gives independent knowledge,
 on 
the structure of the unphysical terms stemming from the 
1PI self-energy contribution of a given order. Instead, we have 
no independent knowledge of the unphysical terms  stemming from the
1PI vertex contribution; the latter may be deduced, if desirable 
through appropriate combination of the  STI 
of $A_{\mu}^a\, A_{\nu}^b\, q^i\, \bar{q}^j$ and the STI of 
the (full) three-gluon vertex mentioned above.
Roughly speaking, the unphysical contributions from the self-energy,
which are known from the latter STI, 
must be canceled against the 
(unknown) unphysical contributions stemming from the vertex, since 
there are no unphysical contributions in the the  STI of  
$A_{\mu}^a\, A_{\nu}^b\, q^i\, \bar{q}^j$, which is 
the sum of the two terms (as we will see in a moment,
a minor refinement to this argument is 
necessary in order to account for 1PR contributions, but the 
general idea is essentially this).

In
particular, denoting by $\g_{A_\alpha A_\mu A_\nu}(q,k_1,k_2)$ the
all order gluon three-point function [with $\g_{A_\alpha A_\mu
A_\nu}^{[0]}\equiv\Gamma^{[0]}_{\alpha\mu\nu}$ as defined in Eq.(\ref{tgv})] 
the 
STI triggered is \cite{Ball:ax}
\bea
k_1^\mu\g_{A_\alpha A_\mu A_\nu}(q,k_1,k_2) & = & 
\left[i\Delta^{(-1)\,\rho}_{\nu}(k_2)+k_{2}^{\rho}k_{2\nu}\right]
\left[k_1^2D(k_1)\right]H_{\rho\alpha}(k_2,q) \nonumber \\
& - & \left[i\Delta^{(-1)\,\rho}_{\alpha}(q)+q^\rho q_\alpha\right]
\left[k_1^2D(k_1)\right]H_{\rho\nu}(q,k_2), \nonumber \\
k_2^\nu\g_{A_\alpha A_\mu A_\nu}(q,k_1,k_2) & = & 
\left[i\Delta^{(-1)\,\rho}_{\alpha}(q)+q^\rho q_\alpha\right]
\left[k_2^2D(k_2)\right]H_{\rho\mu}(q,k_1) \nonumber \\
& - & \left[i\Delta^{(-1)\,\rho}_{\mu}(k_1)+p_{1}^{\rho}k_{1\mu}\right]
\left[k_2^2D(k_2)\right]H_{\rho\alpha}(k_1,q), 
\label{STIbc}
\eea  
where $H$ represents the ghost Green's function appearing in the
conventional formalism (see for example \cite{PasTar}); at tree level
\be
\bpi(0,50)(45,-20)

\PhotonArc(40,0)(20,40,180){-1.5}{6.5}
\DashCArc(40,0)(20,180,320){1}
\DashArrowLine(40.5,-20)(39.5,-20){1}
\Vertex(20,0){1.8}

\Text(10,-1)[r]{$H_{\alpha\beta}^{[0]}(k_1,k_2)\,=$}
\Text(25,0)[l]{\footnotesize{$k_{1\alpha}$}}
\Text(59,13)[l]{\footnotesize{$k_{2\beta}$}}
\Text(120,-1)[r]{$=\,-igg_{\alpha\beta}.$}

\epi
\label{Htl}
\ee 

On the other hand, with the help of the Batalin-Vilkovisky formalism
\cite{Batalin:pb} formulated in the BFG,
 one can relate the BFG gluon two-point function
$\widetilde\g_{\widetilde A_\alpha \widetilde A_\beta}$ with the
conventional one 
$\g_{A_\alpha A_\beta}$ through a ``background-quantum identity''
(BQI) \cite{MFPAG} of the form 
\be
\widetilde\g_{\widetilde A_\alpha \widetilde A_\beta}(q)=\g_{A_\alpha
A_\beta}(q) 
+2\g_{\Omega_\alpha A^*_\rho}(q)\g_{A^\rho A_\beta}(q)+
\g_{\Omega_\alpha A^*_\rho}(q)\g_{A^\rho A^\sigma}(q)\g_{\Omega_\beta
A^*_\sigma}(q), 
\label{BQI}
\ee
where
\be
\g_{{A^a_\alpha}{A^b_\beta}}(q)=\delta^{ab}
\left[iq_\alpha q_\beta-\Delta^{(-1)}_{\alpha\beta}(q)\right] \qquad
\Longrightarrow \qquad
\left\{\begin{array}{l}
\g_{A_\alpha A_\beta}^{[0]}(q)=
-iq^2P_{\alpha\beta}(q), \\ 
\g_{A_\alpha A_\beta}^{[n]}(q)=\Pi^{[n]}_{\alpha\beta}(q^2),
\label{idena}
\end{array}\right.
\ee
and $\g_{\Omega A^*}$ represents an auxiliary (unphysical) two-point
function connecting a background source $\Omega$ with a gluon
anti-field $A^*$ (see \cite{Binosi:2002ez} for details).   

The observation made in \cite{Binosi:2002ez} was that,
even though the auxiliary Green's function appearing in the STI of
Eq.(\ref{STIbc})
is different from the one appearing in the BQI of Eq.(\ref{BQI}), the two
are related by a Schwinger-Dyson type of relation, which reads
\be
i\g_{\Omega_\alpha A^*_\beta}(q)=C_A 
\int
H^{[0]}_{\alpha\rho}(q,-k)
D(k-q)\Delta^{\rho\sigma}(k)H_{\beta\sigma}(-q,k),
\label{gpert2}
\ee
where $C_A$ denotes the Casimir eigenvalue of the adjoint
representation, {\it i.e.}, $C_A=N$ for $SU(N)$, and 
the integral measure is defined in Eq.(\ref{im}). 
Diagrammatically, Eq.(\ref{gpert2}) reads
\be
\bpi(0,60)(20,-25)

\PhotonArc(40,0)(20,0,180){-1.5}{8.5}
\DashCArc(40,0)(20,180,360){1}
\Photon(60,0)(90,0){1.5}{4}
\GCirc(60,0){10}{0.8}
\GCirc(40,20){8}{0.8}
\GCirc(40,-20){8}{0.8}
\DashArrowLine(25.86,-14.14)(25.96,-14.24){1}
\DashArrowLine(54.14,-14.14)(54.24,-14.04){1}
\Vertex(20,0){1.8}
\Text(25,0)[l]{$\scriptstyle{\alpha}$}

\Text(61,0)[c]{$\scriptstyle{H_{\beta\sigma}}$}
\Text(40,-20)[c]{$\scriptstyle{D}$}
\Text(40.5,20)[c]{$\scriptstyle{\Delta^{\!\rho\sigma}}$}
\Text(5,-1)[r]{$i\g_{\Omega_\alpha A^*_\beta}(q)\,=$}
\epi
\ee

Evidently these last equations expresses the two-point 
Green's function
$\g_{\Omega_\alpha A^*_\beta}(q)$, which is definable in the
BV framework, entirely in terms of Green's functions definable in the
conventional formalism; this will in turn connect 
the STI of
Eq.(\ref{STIbc}) and the BQI of Eq.(\ref{BQI}), which is what will finally
allow to prove the 
correspondence between the PT an the BFG to all orders, using the
intrinsic PT algorithm. 
 
\begin{figure}[t]
\bce
\includegraphics[width=15cm]{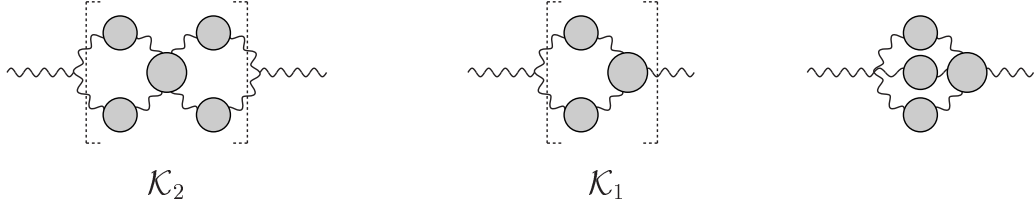}
\ece
\caption{\label{figH} Schematic representation of some 1PI diagrams,
  with their associated kernels, contributing to the all-order gluon
  self-energy.} 
\end{figure}

Following the two-loop case described in \cite{Binosi:2002ez}, 
we will now generalize the intrinsic PT procedure to all orders. 
The  1PI Feynman  diagrams contributing  to the  conventional 
gluon self-energy in the $R_\xi$  gauges can be always 
separated into three distinct
sets (Fig.\ref{figH}):  ({\it  i})   the  set  of  diagrams  that
have  two  external 
(tree-level)   three-gluon   vertices,  and   thus   can  be   written
schematically  (suppressing  Lorentz  indices) as  $\Gamma^{[0]}[{\cal
K}_2]\Gamma^{[0]}$, where ${\cal K}_2$ is some kernel; ({\it ii}) the set
of diagrams  with only  one external (tree-level)  three-gluon vertex,
and  thus  can  be  written  as $\Gamma^{[0]}[{\cal  K}_1]$  or  $[{\cal
K}_1]\Gamma^{[0]}$;  ({\it iii}) all  remaining diagrams,  containing no
external three-gluon vertices.

At this point we make
the following observation: if we carry out  the
decomposition presented in Eq.(\ref{INPTDEC1}) 
to the pair of external  vertices appearing in
the diagrams of the set ({\it i}), and the decomposition of
Eq.(\ref{decomp}) 
to the
external vertex appearing in the diagrams of the set ({\it ii}), 
after a judicious
rearrangement of the kernels ${\cal K}_2$ and ${\cal K}_1$ (together with their
statistical factors), relabeling of internal 
momenta (the  momenta $k_1^\alpha$  and $k_2^\beta$
appearing in Eq.(\ref{STIbc}) will be in fact now related to
virtual  integration  momenta appearing  in  the  quantum loops), and
taking into account the transversality of the gluon self-energy, 
we will end up with the result
\be
\bpi(0,50)(40,-20)

\Text(-70,0)[r]{$\left\{\g_{AA}\right\}^{\rm
P}=$} 

\PhotonArc(0,0)(20,-5,355){1.5}{20}
\Photon(21.5,0)(49.5,0){1.5}{4}
\Photon(-46.5,0)(-21.5,0){1.5}{4}

\Vertex(-21.5,0){1.8}
\Text(-24.5,-8)[r]{$\scriptstyle{\Gamma^{\rm P}}$}

\GCirc(20,0){10}{0.8}
\Text(20.5,0)[c]{$\scriptstyle{\g_{\scriptscriptstyle{\!A\!A\!A}}}$}

\Text(68,0)[c]{$+\ 2$}

\PhotonArc(130,0)(20,-5,355){1.5}{20}
\Photon(151.5,0)(179.5,0){1.5}{4}
\Photon(83.5,0)(108.5,0){1.5}{4}

\Vertex(108.5,0){1.8}
\Text(105.5,-8)[r]{$\scriptstyle{\Gamma^{\rm P}}$}

\GCirc(150,0){10}{0.8}
\Text(150.5,0)[c]{$\scriptstyle{\g_{\scriptscriptstyle{\!A\!A\!A}}}$}

\GCirc(130,-20){8}{0.8}
\Text(130,-20)[c]{$\scriptstyle{\Delta}$}

\epi
\label{alldiag}
\ee

Thus, the longitudinal terms $k_{1}^{\alpha}$ and $k_{2}^{\beta}$
stemming from  $\Gamma^{{\rm P}}_{\alpha \mu \nu}(q,k_1,k_2)$  
will be triggering the STIs of Eq.(\ref{STIbc}). For example, at the $n$-loop
level, one would be triggering the
($m$)-loop version (with $m=0,1,\dots,n-1$) of
the aforementioned STIs.
Therefore,  the all order generalization of the
intrinsic PT would amount to  isolating from 
Eq.(\ref{alldiag}) the terms of the STI of Eq.(\ref{STIbc}) 
that are proportional to 
$[\Delta^{(-1)\,\rho}_{\alpha}(q)]$
($[\Delta^{(-1)\,\rho}_{\alpha}(q)]^{[m]}$ in the $n$-loop
case); we will 
denote such contributions by  $\Pi_{\alpha\beta}^{{\rm
IP}}(q)$. Thus the 1PI diagrams contributing to the
gluon self-energy can be cast in the form
\be
\g_{A_\alpha A_\beta}(q)=G_{A_\alpha A_\beta}(q)+
\Pi_{\alpha\beta}^{{\rm IP}}(q).
\label{ip1}
\ee 

Notice however that   
the 1PR set ${\mathbb S}$ containing  diagrams constructed from
strings of lower order self-energy graphs (the set ${\mathbb
S}^{[n]}$ containing the $2^{n-1}$ diagrams constructed from
strings of self-energy insertions of order less than $n$, in our $n$-loop
example),    
must also be rearranged following the intrinsic PT procedure, and 
converted into the equivalent set $\widehat{\mathbb S}$ containing 
strings involving PT self-energies. 
This treatment of the 1PR strings will
give rise, in addition to the PT strings, 
to ({\it i}) a set of contributions which are proportional to the
inverse tree-level propagator of the external legs $\de(q)$ (with
$d(q)=-i/q^2$ the RFG  
tree-level gluon propagator), and ({\it ii})
a set of contributions which is {\it
effectively} 1PI, and therefore  also belongs to the definition of the 1PI
PT gluon self-energy; we will denote these two sets of
contributions 
collectively by $S^{{\rm IP}}_{\alpha\beta}(q)$. Thus the sum of 
the 1PI and 1PR contributions to the conventional gluon  
self-energy can be cast in the form  
\be
\g_{A_\alpha A_\beta}(q)+{\mathbb S}_{\alpha\beta}(q)
=G_{A_\alpha A_\beta}(q)+\widehat{\mathbb S}_{\alpha\beta}(q)+
\Pi_{\alpha\beta}^{{\rm IP}}(q)+S^{{\rm
IP}}_{\alpha\beta}(q).
\label{ip2}
\ee

By definition of the intrinsic PT procedure, we will now discard from the above
expression all the terms which are proportional to the inverse
propagator of the external legs, 
thus defining the quantity
\be
R^{{\rm
IP}}_{\alpha\beta}(q)=\Pi'^{\,{\rm IP}}_{\alpha\beta}(q)+S'^{\,{\rm
IP}}_{\alpha\beta}(q),
\label{ip3}
\ee
where the primed functions are defined starting from the unprimed ones
appearing in Eq.(\ref{ip2}) by discarding the aforementioned terms.

Thus, making use of Eqs.(\ref{ip1}), (\ref{ip2}) and (\ref{ip3}),
the intrinsic PT gluon
self-energy, to be denoted as ${\widehat\g}_{A_\alpha
A_\beta}(q)$, will be finally defined to all orders as
\bea
{\widehat\g}_{A_\alpha A_\beta}(q)&=& G_{A_\alpha
A_\beta}(q)+R^{{\rm IP}}_{\alpha\beta}(q)\nonumber \\
&=&\g_{A_\alpha A_\beta}(q) - \Pi_{\alpha\beta}^{{\rm IP}}(q) +
R^{{\rm IP}}_{\alpha\beta}(q).
\label{2lPTtwopfIP}
\eea

We next proceed to the construction of the quantities
$\Pi_{\alpha\beta}^{{\rm IP}}(q)$ and $R^{{\rm IP}}_{\alpha\beta}(q)$
discussed above.

\subsection{1PI diagrams}

From Eq.(\ref{alldiag}) and the transversality of the gluon propagator, we
find that the pinching contributions coming from the 1PI diagrams can
be written as
\be
\left\{\g_{A_\alpha A_\beta}\right\}^{\rm
P}=- 2C_A\int\! d(k) \Gamma^{\rm
P}_{\alpha\mu\nu}(q,k-q,-k)\Delta^\nu_\sigma(k) \g_{A_\beta A^\mu
A^\sigma}(q,k-q,-k). 
\ee
Using then the definition of $\Gamma^{\rm P}$ given in Eq.(\ref{GFGP})
together with the tree-level value of the $H$ Green's function [see
Eq.(\ref{Htl})], we get
\be
\left\{\g_{A_\alpha A_\beta}\right\}^{\rm
P}=2iC_A\int\! d(k) k^{\mu}H^{[0]}_{\alpha\nu}(q,-k)
\Delta^{\nu\sigma}(k) \g_{A_\beta A_\mu
A_\sigma}(q,k-q,-k).
\ee

To construct the PT quantity $\Pi^{\rm IP}_{\alpha\beta}(q)$, we now
use the STI of Eq.(\ref{STIbc}) (with $k_1=k-q$ and $k_2=-k$) keeping only
pinching terms; with the help of Eq.(\ref{idena}) we then find 
\be
\Pi^{\rm
IP}_{\alpha\beta}(q)=2iC_A\int\!H^{[0]}_{\alpha\nu}(q,-k)
D(k-q) \Delta^{\nu\sigma}(k)H_{\rho\sigma}(-q,k)\g_{A^\rho
A_\beta}(q),
\ee
which, using Eq.(\ref{gpert2}), can be finally cast in the form
\be
\Pi^{\rm
IP}_{\alpha\beta}(q)=-2\g_{\Omega_\alpha A^*_\rho}(q)\g_{A^\rho
A_\beta}(q).
\label{IPPI}
\ee

\subsection{1PR diagrams}

From the 1PR set of diagrams ${\mathbb S}_{\alpha\beta}$, we need to
identify the subset of contributions $S'^{\, \rm{IP}}_{\alpha\beta}$
which is effectively 1PI. In what follows, to avoid notational clutter
we will suppress Lorentz indices.

The key observation for constructing the aforementioned quantity
$S'^{\, \rm{IP}}$, is that at any order the only elements of the 1PR
set $\mathbb S$ that can contribute to it are the strings that
contains at most three self-energy insertions, {\it i.e.}, the subsets
\bea
{\mathbb S}_2&=&\g_{AA}\,d\,\g_{AA}, \nonumber  \\
{\mathbb S}_3&=&\g_{AA}\,d\,\g_{AA}\,d\,\g_{AA}.
\eea

To understand the reason for that, let us consider the order $n$ set of
1PR diagrams ${\mathbb S}^{[n]}$, and suppose that the PT construction
has been successfully carried out at order $n-1$.
Then consider a generic string ${\mathbb S}^{[n]}_m\subset{\mathbb
S}^{[n]}$ which contains $m$ self-energy insertions
\be
{\mathbb S}^{[n]}_m=\g_{AA}^{[i_1]}\,d\,\g_{AA}^{[i_2]}\,d\,\cdots
\,d\,\g_{AA}^{[i_\ell]}\,d\,\cdots\,
d\,\g_{AA}^{[i_{m-1}]}\,d\,\g_{AA}^{[i_m]}, 
\label{IP1}
\ee
where $\sum_{k=1}^mi_k\equiv n$.

We now concentrate on the self-energy insertion $\g_{AA}^{[i_\ell]}$
appearing in Eq.(\ref{IP1}), and convert it into a PT
self-energy insertion $\widehat\g^{[i_\ell]}_{AA}$. Since $i_\ell<n$,
we know that
$\widehat\g^{[i_\ell]}_{AA}\equiv\widetilde\g^{[i_\ell]}_{\widetilde 
A\widetilde A}$, and we can use the BQI of Eq.(\ref{BQI}) to relate the
BFG self-energy
$\widetilde\g^{[i_\ell]}_{\widetilde A\widetilde A}$ to the conventional one
$\g_{AA}^{[i_\ell]}$. Thus from the aforementioned PT conversion 
one will get the following extra terms
\bea
& & \g_{AA}^{[i_1]}\,d\,\g_{AA}^{[i_2]}\,d\,\cdots\,d\,
\left(\g_{AA}^{[i_{\ell-1}]}\g_{\Omega A^*}^{[i_\ell]}\right)\,d\,
\cdots\, d\,\g_{AA}^{[i_{m-1}]}\,d\,\g_{AA}^{[i_m]}, \nonumber  \\
& &\g_{AA}^{[i_1]}\,d\,\g_{AA}^{[i_2]}\,d\,\cdots\,d\,
\left(\g_{\Omega A^*}^{[i_\ell]}\g_{AA}^{[i_{\ell+1}]}\right)
\,d\,
\cdots\, d\,\g_{AA}^{[i_{m-1}]}\,d\,\g_{AA}^{[i_m]}, \nonumber  \\
& & \g_{AA}^{[i_1]}\,d\,\g_{AA}^{[i_2]}\,d\,\cdots\,d\,
\left(-2\sum_{j=1}^{i_\ell-1}\g_{\Omega A^*}^{[i_\ell-j]} \g_{AA}^{[j]}
\right)
\,d\,
\cdots\, d\,\g_{AA}^{[i_{m-1}]}\,d\,\g_{AA}^{[i_m]},\nonumber \\
& & \g_{AA}^{[i_1]}\,d\,\g_{AA}^{[i_2]}\,d\,\cdots\,d\,
\left(-\sum_{j_1=1}^{i_\ell-1}\sum_{j_2=0}^{j_1-1}
\g_{\Omega A^*}^{[i_\ell-j_1]} \g_{AA}^{[j_2]}\g_{\Omega A^*}^{[j_1-j_2]}
\right)
\,d\,
\cdots\, d\,\g_{AA}^{[i_{m-1}]}\,d\,\g_{AA}^{[i_m]}.
\label{extrastuff}
\eea
The first two comes from the part of the BQI of Eq.(\ref{BQI})
proportional to $d^{-1}(q)$ and will contribute to cancel
the terms one has to add in the conversion 
to the following two strings
of the subset ${\mathbb S}^{[n]}_{m-1}$:
\bea
& &\g_{AA}^{[i_1]}\,d\,\g_{AA}^{[i_2]}\,d\,\cdots\,d\,\g_{AA}^{[i_{\ell-2}]}\,
d\,\g^{[i_{\ell-1}+i_\ell]}_{AA}\,d\,\cdots\,d\,
\g_{AA}^{[i_{m-2}]}\,d\,\g_{AA}^{[i_{m-1}]},\nonumber \\
& &\g_{AA}^{[i_1]}\,d\,\g_{AA}^{[i_2]}\,d\,\cdots\,d\,\g_{AA}^{[i_{\ell-1}]}\,
d\,\g^{[i_\ell+i_{\ell+1}]}_{AA}\,d\,\cdots\,d\,
\g_{AA}^{[i_{m-2}]}\,d\,\g_{AA}^{[i_{m-1}]}.
\eea

The last two terms in (\ref{extrastuff}) will precisely cancel
the terms leftover from the PT conversion of the string ${\mathbb
S}^{[i_\ell]}$ appearing in the following subset of ${\mathbb S}^{[n]}_m$
\be
\g_{AA}^{[i_1]}\,d\,\g_{AA}^{[i_2]}\,d\,\cdots\,d\,{\mathbb
S}^{[i_\ell]} \,d\,\cdots\,d\g_{AA}^{[i_{m-1}]}\,d\,\g_{AA}^{[i_m]}.
\ee

We therefore see that the terms that one needs to add to a string of
order $n$, which contains more than three self-energy insertions, will be
canceled by other strings of the same order, but containing a
different number of insertions.
The only time that one will obtain terms that do not cancel and, as such,
must be added to the 1PI gluon two-point function $\g^{[n]}_{A_\alpha
A_\beta}$, is if the string contain two or three self-energy insertions
($\mathbb S^{[n]}_2$ and $\mathbb S^{[n]}_3$ respectively). In
this case we will get
\bea
{\mathbb S}^{[n]}_2&\to&\widehat{\mathbb S}^{[n]}_2+
2\sum_{m=1}^{n-1}\g_{\Omega A^*}^{[n-m]}\g_{AA}^{[m]}+\sum_{m=1}^{n-1}
\g_{\Omega A^*}^{[n-m]}\g_{AA}^{[0]}\g_{\Omega A^*}^{[m]}+4
\sum_{m=2}^{n-1}\sum_{\ell=1}^{m-1}
\g_{\Omega A^*}^{[n-m]}\g_{AA}^{[\ell]}\g_{\Omega A^*}^{[m-\ell]},
\nonumber \\
{\mathbb S}^{[n]}_3&\to&\widehat{\mathbb S}^{[n]}_3-
3\sum_{m=2}^{n-1}\sum_{\ell=1}^{m-1}
\g_{\Omega A^*}^{[n-m]}\g_{AA}^{[\ell]}\g_{\Omega A^*}^{[m-\ell]}.
\eea

Thus we see that the total effective 1PI 
contribution coming from the conversion of
the $2^{n-1}$ 1PR strings contributing to the gluon self-energy
$\g^{[n]}_{A_\alpha A_\beta}$ at order $n$, into the corresponding 
1PR PT strings, will be
\be
S'^{\,{\rm IP}\,[n]}=2\sum_{m=1}^{n-1}\g_{\Omega A^*}^{[n-m]}\g_{AA}^{[m]}+
\sum_{m=1}^{n-1}\sum_{\ell=0}^{m-1}
\g_{\Omega A^*}^{[n-m]}\g_{AA}^{[\ell]}\g_{\Omega A^*}^{[m-\ell]}.
\ee
On the other hand, Eq.(\ref{IPPI}) implies that
\be
\Pi'^{\,{\rm
IP}\,[n]}=-2\sum_{m=1}^{n-1}
\g_{\Omega A^*}^{[n-m]}\g^{[m]}_{AA},
\ee
so that adding by parts the last two equations and 
putting back Lorentz and momentum indices, 
we get the all order result
\be
R^{{\rm IP}}_{\alpha\beta}(q)=
\g_{\Omega_\alpha A^*_\mu}(q)\g_{A^\mu A^\nu}(q)
\g_{\Omega_\beta A^*_\nu}(q).
\ee
Thus, making use of the BQI of Eq.(\ref{BQI}), we have the identity
\bea
{\widehat\g}_{A_\alpha A_\beta}(q)&=&
\g_{A_\alpha A_\beta}(q) - \Pi_{\alpha\beta}^{{\rm IP}}(q) +
R^{{\rm IP}}_{\alpha\beta}(q)\nonumber \\
&=&\g_{A_\alpha A_\beta}(q)+
2\g_{\Omega_\alpha A^*_\rho}(q)\g_{A^\rho
A_\beta}(q)
+\g_{\Omega_\alpha A^*_\rho}(q)\g_{A^\rho A^\sigma}(q)
\g_{\Omega_\beta A^*_\sigma}(q) \nonumber \\
&=&\widetilde \g_{\widetilde A_\alpha \widetilde A_\beta}(q),
\eea
a result that completes the all-order explicit construction of the 
PT gluon self-energy. 

\section{Process independence of the PT algorithm} 

One important question to be addressed in the PT context, is whether this 
construction depends on the specific kind of
external particles chosen. 
This issue was addressed in 
\cite{Watson:1994tn}  by
means of detailed calculations, and in
\cite{Binosi:2002ez} through
the use of the BQIs.
In both
cases it was shown that, at the one-loop 
level, the gluon self-energy constructed by resorting to the
$S$-matrix PT algorithm is universal, in the sense that its form 
does not depend on the specific process used for the embedding.

The purpose of this section is to demonstrate that this property holds true
to all orders. Before proving this in the most general case, let us
consider a specific example, {\it i.e.}, the construction of the PT
gluon self-energy through the process
$A^{d_1}_{\rho_1}(r_1)A^{d_2}_{\rho_2}(r_2)\to
A^{e_1}_{\sigma_1}(p_1)A^{e_2}_{\sigma_2}(p_2)$, where
$A^{d_i}_{\rho_i}(r_i)$ and $A^{e_i}_{\sigma_i}(p_i)$ represent on-shell
gluons, {\it i.e.}, with $r_i^2=p_i^2=0$ and
$r_i^{\rho_i}\epsilon_{\rho_i}(r_i)=p_i^{\sigma_i}\epsilon_{\sigma_i}(p_i)=0$. 

As before we will denote by 
${\cal A}^{d_1d_2e_1e_2}(r_1,r_2,p_1,p_2)$ the subset of graphs which
will receive the action of the longitudinal momenta stemming from the
pinch part $\Gamma^{{\rm P}}_{\alpha\mu\nu}(q,k_1,k_2)$
of the three-gluon vertex. We have then
\bea
{\cal A}^{d_1d_2e_1e_2}(r_1,r_2,p_1,p_2)&=&g\epsilon^{\rho_1}(r_1)
\epsilon^{\rho_2}(r_2)\Gamma^{ed_1d_2}_{\alpha\rho_1\rho_2}(q,r_1,r_2)f^{eab}
\Gamma^{{\rm P},\alpha\mu\nu}(q,k_1,k_2)\times \nonumber \\
&\times&{\cal T}^{abe_1e_2}_{\mu\nu}(k_1,k_2,p_1,p_2), 
\eea
where now ${\cal T}^{abe_1e_2}_{\mu\nu}$ represents the sub-amplitude
$g^{a}_{\mu}(k_1)g^{b}_{\nu}(k_2)\to
g^{e_1}_{\sigma_1}(p_1)g^{e_2}_{\sigma_2}(p_2)$ with the initial
gluons off-shell and final ones on-shell. Diagrammatically then,
\bce
\bpi(0,60)(0,-30)

\Text(-47,-7)[r]{${\cal A}^{abe_1e_2}=$}

\Photon(-15,-7)(-42,20){1.5}{5}

\Photon(-42,-34)(-15,-7){-1.5}{5}

\Text(-45,20)[r]{$\scriptstyle{r_1}$}
\Text(-45,-34)[r]{$\scriptstyle{r_2}$}
\Photon(-15,-7)(1,-7){1.5}{3}
\Text(-15,-13)[l]{$\scriptstyle{\alpha}$}
\Text(-15,-1)[l]{$\scriptstyle{e}$}

\PhotonArc(17.5,-7.5)(15,0,360){1.5}{17}
\GCirc(17.5,6.5){5}{0.8}
\Text(17.5,6.5)[c]{$\scriptstyle{\Delta}$}
\Text(5,7.5)[c]{$\scriptstyle{\nu}$}
\Text(32.5,7.5)[c]{$\scriptstyle{\sigma}$}
\GCirc(17.5,-22.5){5}{0.8}	
\Text(17.5,-22.5)[c]{$\scriptstyle{\Delta}$}
\Text(5,-22.5)[c]{$\scriptstyle{\mu}$}
\Text(31.5,-22.5)[c]{$\scriptstyle{\rho}$}

%\ArrowLine(60,10)(40,-5)
%\ArrowLine(75,20)(60,10)
%\GCirc(58.5,8.5){5}{0.8}
%\Text(58.5,8.5)[c]{$\scriptstyle{S}$}
\Photon(65,20)(38,-7){1.5}{5}
\Text(70,20)[l]{$\scriptstyle{p_1}$}
\Text(57.5,20)[r]{$\scriptstyle{\sigma_1}$}

%\ArrowLine(40,-10)(60,-25)
%\ArrowLine(60,-25)(75,-35)
%\GCirc(58.5,-23.5){5}{0.8}
%\Text(58.5,-23.5)[c]{$\scriptstyle{S}$}
\Photon(38,-7)(65,-34){-1.5}{5}
\Text(70,-34)[l]{$\scriptstyle{p_2}$}
\Text(57.5,-34)[r]{$\scriptstyle{\sigma_2}$}

\GCirc(41.5,-7.5){13}{0.8}
\Text(42,-7.5)[c]{${\cal C}^{\sigma_1\sigma_2}_{\rho\sigma}$} 

\DashLine(8,-39)(8,25){1}
\DashLine(8,25)(11,25){1}
\DashLine(8,-39)(11,-39){1}
\DashLine(80,-39)(80,25){1}
\DashLine(80,25)(77,25){1}
\DashLine(80,-39)(77,-39){1}

\epi
\ece
so that in terms of Green's functions we have 
\be
{\cal T}^{abe_1e_2}_{\mu\nu}(k_1,k_2,p_1,p_2)=\left[
\Delta_\mu^\rho(k_1)\Delta_\nu^\sigma(k_2) {\cal
C}^{abe_1e_2}_{\rho\sigma\sigma_1\sigma_2}(k_1,k_2,p_1,p_2)\right] 
\epsilon^{\sigma_1}(p_1)\epsilon^{\sigma_2}(p_2).
\label{ampli}
\ee
Clearly there is an equal contribution coming from a mirror diagram
where $\Gamma^{\rm P}$ is situated to the right-hand side of the  
${\cal T}^{abe_1e_2}_{\mu\nu}$ amplitude.
As in the quark--anti-quark case, we need to focus on the STI satisfied by the
amplitude (\ref{ampli}). To this end, we start from the trivial identities
\bea
\bigg\langle T\left[\bar c^a(x)A^b_\nu(y)A^{e_1}_{\lambda_1}(z)
A^{e_2}_{\lambda_2}(w)\right]\bigg\rangle&=&0,\nonumber \\
\bigg\langle T\left[A^a_\mu(x)\bar c^b(y)A^{e_1}_{\lambda_1}(z)
A^{e_2}_{\lambda_2}(w)\right]\bigg\rangle&=&0, 
\eea
and then apply the BRST operator $s$ of Eq.(\ref{BRST}), to get the STIs
\bea
& & \partial^\mu_xC^{abe_1e_2}_{\mu\nu\lambda_1\lambda_2}+\partial_\nu^y
G_{1\lambda_1\lambda_2}^{abe_1e_2}+gf^{bcd}
Q_{1\nu\lambda_1\lambda_2}^{acde_1e_2}\nonumber \\
&+& \partial_{\lambda_1}^z
G_{1\nu\lambda_2}^{abe_1e_2}+ \partial_{\lambda_2}^w
G_{1\nu\lambda_1}^{abe_1e_2}+gf^{e_1cd}
Q_{1\nu\lambda_1\lambda_2}^{abcde_2}+gf^{e_2cd}
Q_{1\nu\lambda_1\lambda_2}^{abe_1cd}=0, \nonumber \\
& & \partial^\mu_yC^{abe_1e_2}_{\mu\nu\lambda_1\lambda_2}+\partial_\mu^x
G_{2\lambda_1\lambda_2}^{abe_1e_2}+gf^{acd}
Q_{2\mu\lambda_1\lambda_2}^{cdbe_1e_2}\nonumber \\
&+& \partial_{\lambda_1}^z
G_{2\mu\lambda_2}^{abe_1e_2}+ \partial_{\lambda_2}^w
G_{2\mu\lambda_1}^{abe_1e_2}+gf^{e_1cd}
Q_{2\mu\lambda_1\lambda_2}^{abcde_2}+gf^{e_2cd}
Q_{2\mu\lambda_1\lambda_2}^{abe_1cd}=0,
\label{psSTI1}
\eea
where the Green's functions  
appearing above (in configuration space) are
obtained from the corresponding ones appearing in Eq.(\ref{psgf}), through
the replacements $\bar q^i(z)\to A^{e_1}_{\lambda_1}(z)$ and
$q^j(w)\to A^{e_2}_{\lambda_2}(w)$ plus an eventual suitable
permutation of fields; for example,
\bea
G_{1\nu\lambda_2}^{abe_1e_2}&=&\bigg\langle T\left[\bar
c^a(x)A^b_\nu(y)c^{e_1}(z)  
A^{e_2}_{\lambda_2}(w)\right]\bigg\rangle, \nonumber \\
Q_{2\mu\lambda_1\lambda_2}^{abcde_2}&=&\bigg\langle T\left[A^a_\mu(x)\bar
c^b(y)A^c_{\lambda_1}(z)c^d(z)A^{e_2}_{\lambda_2}(w)\right]\bigg\rangle.
\eea

We can then Fourier transform the identity of
Eq.(\ref{psSTI1}) to obtain the momentum-space STIs
\bea
& & k^\mu_1C^{abe_1e_2}_{\mu\nu\lambda_1\lambda_2}+
k_{2\nu}G_{1\lambda_1\lambda_2}^{abe_1e_2}-igf^{bcd}
Q_{1\nu\lambda_1\lambda_2}^{acde_1e_2} \nonumber \\
&+&p_{1\lambda_1}G_{1\nu\lambda_2}^{abe_1e_2}+p_{2\lambda_2}
G_{1\nu\lambda_1}^{abe_1e_2}-igf^{e_1cd}
Q_{1\nu\lambda_1\lambda_2}^{abcde_2}-igf^{e_2cd}
Q_{1\nu\lambda_1\lambda_2}^{abe_1cd}=0, \nonumber \\
& & k^\nu_2C^{abe_1e_2}_{\mu\nu\lambda_1\lambda_2}+
k_{1\mu}G_{2\lambda_1\lambda_2}^{abe_1e_2}-igf^{acd}
Q_{2\mu\lambda_1\lambda_2}^{cdbe_1e_2} \nonumber \\
&+&p_{1\lambda_1}G_{2\mu\lambda_2}^{abe_1e_2}+p_{2\lambda_2}
G_{2\mu\lambda_1}^{abe_1e_2}-igf^{e_1cd}
Q_{2\mu\lambda_1\lambda_2}^{abcde_2}-igf^{e_2cd}
Q_{2\mu\lambda_1\lambda_2}^{abe_1cd}=0, 
\label{msSTI1}
\eea
where the momentum-space Green's functions appearing above are
obtained from the corresponding ones appearing in Eq.(\ref{msgf}), by
replacing the fermion propagators $S(p_i)$ with the gluon propagators
$\Delta_{\lambda_i}^{\tau_i}(p_i)$, and 
adding the corresponding Lorentz index $\tau_i$ to the kernel involved
in the definition.

The last four terms of both the STIs of Eq.(\ref{msSTI1}) will actually
die due to the on-shell condition of the external gluons. In fact, we
multiply both sides of Eq.(\ref{msSTI1}) by the product 
$\Delta^{(-1)\lambda_1}_{\sigma_1}(p_1)\Delta^{(-1)\lambda_2}_{\sigma_2}(p_2)$
of the two inverse propagators of the external gluons, and then
contract the resulting amplitudes with the polarization tensors
$\epsilon^{\sigma_i}(p_i)$. Since the external gluon are assumed to be
on-shell, we have that
\bea
& & \epsilon^{\sigma_i}(p_i) \Delta^{(-1)\lambda_i}_{\sigma_i}(p_i)=0,
\nonumber \\
& &  \epsilon^{\sigma_i}(p_i)p_{i\sigma_i}=0,
\eea
from which the vanishing of the aforementioned terms follows.
Thus we arrive at the on-shell STIs
\bea
k_1^\mu{\cal T}_{\mu\nu}^{abe_1e_2}(k_1,k_2,p_1,p_2)&=&
{\cal S}_{1\nu}^{abe_1e_2}(k_1,k_2,p_1,p_2), \nonumber \\
k_2^\nu{\cal T}_{\mu\nu}^{abe_1e_2}(k_1,k_2,p_1,p_2)&=&
{\cal S}_{2\mu}^{abe_1e_2}(k_1,k_2,p_1,p_2),
\label{onshSTI1}
\eea
with
\bea
{\cal S}_{1\nu}^{abe_1e_2}(k_1,k_2,p_1,p_2)&=&\left[
igf^{bcd}{\cal Q}_{1\nu\sigma_1\sigma_2}^{acde_1e_2}
(k_1,k_2,p_1,p_2)D(k_1)\right.\nonumber \\
&-&\left. 
k_{2\nu}G_{1\sigma_1\sigma_2}^{abe_1e_2}(k_1,k_2,p_1,p_2)D(k_1)D(k_2)
\right]\epsilon^{\sigma_1}(p_1)\epsilon^{\sigma_2}(p_2), \nonumber \\
{\cal S}_{2\mu}^{abe_1e_2}(k_1,k_2,p_1,p_2)&=&\left[
igf^{acd}
{\cal Q}_{2\mu\sigma_1\sigma_2}^{cdbe_1e_2}
(k_1,k_2,p_1,p_2)D(k_2)\right.\nonumber \\
&-&\left. 
k_{1\mu}G_{2\sigma_1\sigma_2}^{abe_1e_2}(k_1,k_2,p_1,p_2)D(k_1)D(k_2)
\right]\epsilon^{\sigma_1}(p_1)\epsilon^{\sigma_2}(p_2).
\eea

These STIs have exactly the same form of the ones shown in
Eq.(\ref{onshSTI}) derived in the quark--anti-quark case. The only
difference is in the kernels which enters in the definitions of the
amplitude ${\cal T}$ and the Green's functions ${\cal S}$. However the
all orders PT algorithm constructed in the previous sections does not
depend in any way on the kernels involved, so that it goes through
unmodified also in the present case. 

Notice that the reason for which the STIs of Eqs.(\ref{onshSTI}) and
(\ref{onshSTI1}) have the same form, is due to the fact that 
the BRST variation of an on-shell field (independently of it being a
quark or a gluon)
vanishes due to the on-shell
condition. Thus, the particular STIs needed for the application of the PT
algorithm are completely determined by the off-shell particles, which
are fixed ({\it i.e.}, two gluons), 
regardless of the process in which
we embed the two-point function we want to construct. 

We thus conclude the construction of 
the PT gluon self-energy through the  embedding into the process
$A^{d_1}_{\rho_1}(r_1) A^{d_2}_{\rho_2}(r_2) A^{d_3}_{\rho_3}(r_3)\to 
A^{e_1}_{\sigma_1}(p_1) A^{e_2}_{\sigma_2}(p_2)
A^{e_3}_{\sigma_3}(p_3)$ with on-shell initial and final gluons, 
proceeds in exactly the same way as in the case of final on-shell quarks. 
The only  adjustments required are those pertaining to the 
kernels appearing in the corresponding STIs, while the construction 
algorithm itself remains unaltered. 

\section{Renormalization}

In this section we will discuss the renormalization of 
the PT Green's functions constructed in the previous sections. 
There is of course no doubt that if one supplies
the correct counterterms within the conventional 
formulation, the entire $S$-matrix will continue being renormalized,
even after the PT rearrangement of the (unrenormalized)
Feynman graphs.
The question addressed in this section is  
whether the new Green's function
constructed through the PT rearrangement are {\it individually}
renormalizable \cite{unitary}. 

The general methodology for dealing with this issue has been established
in the second paper of \cite{Papavassiliou:2000az}, where the two-loop
case was studied in detail: One should start out with the counterterms
which  are  necessary  to  renormalize individually  the  conventional
Green's functions contributing to  the $n$-loop $S$-matrix in the RFG.
Then, one should show  that, by simply rearranging these counterterms,
following the PT rules, one arrive at renormalized $n$-loop PT Green's
functions.  This  section is meant  to serve as a general  framework 
for the all-order construction, 
putting particular emphasis on the various
conceptual  and   methodological  issues  involved,   rather  than  an
explicit proof  of renormalizability.  We  consider  this
discussion sufficient  for convincing the  reader that renormalization
poses  no problem  whatsoever to  the all-order  PT  construction.  
The basic points are the following:

({\it i}) We will assume that the massless Yang-Mills theory, quantized in   
the RFG, is renormalizable to 
all-orders.
We will use the following notation:
$Z_1$ is the vertex renormalization constant for the
quark-gluon vertex $\Gamma_{\alpha}$, 
$Z_2$ is the wave-function renormalization for the (external) quarks,
$Z_A$ the gluon wave-function renormalization
corresponding to the gluon self-energy $\Pi$,
$Z_3$ is the vertex renormalization constant for the 
three-gluon vertex $\Gamma_{\alpha\mu\nu}$, 
${\bar Z}_2$ is the usual ghost wave-function renormalization,
and ${\bar Z}_1$ the ghost-gluon vertex renormalization 
constant; of course, all above quantities and renormalization constants 
are to be computed in the RFG.
Notice also that, the BRST symmetry demands that
$Z_3/{Z_A}  = \bar{Z}_1/\bar{Z}_2$.
Equivalently, one can carry out the renormalization program
using appropriately defined counter-terms.
The corresponding counterterms, which, when added to the
above $n$-loop quantities 
render them UV finite, are, respectively
$K_1^{[n]}$, $K_2^{[n]}$, $K_A^{[n]}$,
$\widehat K_A^{[n]}$,
$K_3^{[n]}$, $K_{3F}^{[n]}$,
$\bar{K}_2^{[n]}$, and $\bar{K}_1^{[n]}$. 
The $Z$'s and the $K$'s are in general related by
$
Z = 1+ \sum_{j=1} K^{[j]}$. 
Of course, mass counterterms $\delta m$  must also be supplied if 
the quarks are considered to be massive.

({\it ii}) It is important to
recognize that, even though the PT self-energies 
does not coincide with the ones appearing inside the 
loops (exactly as happens in the background field method)
there is no conflict with renormalization. This point is rather subtle,
and deserves some further clarification.
At the level of the original Lagrangian (in the RFG) 
the counterterms will be furnished as usual, {\it i.e.} in such a way 
as to render the self-energy and vertices finite. At one-loop,
for example,   
 a counterterm of the form 
$(q^2 g_{\mu\nu} - q_{\mu}q_{\nu}) K_A^{[1]}$ must be provided 
to the self-energy $\Pi^{[1]}$, 
and a term $K_1^{[1]}(\lambda/2)\gamma_{\alpha}$
to the vertex $\Gamma_{\alpha}^{[1]}$. 
The PT self-energy $\widehat{\Pi}^{[1]}$ and 
vertex $\widehat{\Gamma}_{\alpha}^{[1]}$ have different renormalization 
properties than $\Pi^{[1]}$ and  $\Gamma_{\alpha}^{[1]}$; 
therefore, 
the existing counterterms 
must be appropriately reshuffled.
In particular, due to the fact that, unlike $\Gamma_{\alpha}^{[1]}$, 
the PT vertex satisfies the QED-like WI of Eq.(\ref{GS})  
it becomes
ultraviolet finite when the counterterms $K_2^{[1]}$, 
equal to that of the (external) quarks,  
is added to it. Consequently, the remaining 
vertex counterterm, {\it i.e.} $K_1^{[1]}-K_2^{[1]} $, together 
with an equal contribution from the mirror-vertex, 
must be given to $\widehat{\Pi}^{[1]}$; this is accomplished 
by inserting, as usual, 
the unity as $q^{2}(1/q^{2})$ and adding the missing 
longitudinal pieces for free. 
Thus, the resulting (effective) counterterm for $\hat{\Pi}^{[1]}$ will 
be $\widehat{K}^{[1]}_A = K^{[1]}_A - 2 (K_1^{[1]}-K_2^{[1]})$,
and is, of course, equal to the counterterm necessary to 
renormalize 
$\widetilde \g_{\widetilde A_\alpha \widetilde A_\beta}^{[1]}$. 
At this point $\widehat{K}^{[1]}_A$ can effectively be thought 
off as a new propagator-like counterterm. Of course, 
exactly as happens in the BFG, when 
going to the next order the counterterm alloted to $\Pi^{[1]}$ 
appearing inside loops will still be  $K^{[1]}_A$ and not 
$\widehat{K}^{[1]}_A$, {\it i.e.} one must start out, at any given order, 
with the counterterms generated by the 
original Lagrangian defined in the RFG, and rearrange them appropriately. 
Notice also that, again due to the validity of Eq.(\ref{GS}), 
the renormalization constants 
before and after the PT rearrangements 
are related to the 
gauge coupling renormalization as follows:  
\bea
Z^2_g &=& Z_1^2 Z_{2}^{-2} Z_A^{-1} 
=   \widehat{Z}_1^{2} \widehat{Z}_2^{-2}\widehat{Z}_A^{-1}
= \widehat{Z}_A^{-1}.
\label{R3}
\eea

({\it iii}) Of course, 
primitively divergent 
graphs which are inert under the PT rearrangement, such as 
the third graph of Fig.8, 
are rendered finite when their usual 
counterterms are furnished,
without any need for further modifications. 
The same is true for the entire PT box, since it coincides 
with the conventional box in the RFG (and the BFG); therefore 
it has no primitive divergence, and all its sub-divergences 
are canceled by the normal counterterms. 

({\it iv}) The bare three-gluon vertices 
$\Gamma^{eab\,[0]}_{\alpha\mu\nu}$ associated to counter-terms 
do {\it not} undergo the PT splitting of Eq.(\ref{decomp}).
This is consistent with the general PT rules, simply because 
such terms are essentially furnished in order to cancel 
divergences stemming from sub-integrations; as we have explained 
earlier, longitudinal pieces induced by sub-integrations should 
{\it not} pinch, in order not to violate the manifest analyticity 
of the individual Green's functions. The simplest way to see that, 
once pinching induced by sub-integration has been forbidden, 
the counterterms proportional to $\Gamma^{eab\,[0]}_{\alpha\mu\nu}$ 
should not pinch either, is to consider the first 
one-loop vertex diagram 
appearing on the second row of Fig.7 (denoted by 
$[{\cal T}^{[1]}]_{s,\scriptscriptstyle{\rm I}}$), 
and imagine that 
the gluonic triangle has been replaced by a fermionic one.
Evidently 
the resulting graph cannot furnish pinching momenta; on the other hand, its 
divergent part is proportional to $\Gamma^{eab\,[0]}_{\alpha\mu\nu}$,
and so is the counterterm which must be supplied to render it finite.
Clearly, splitting the counterterm, while the main digram is inert, 
will result in an obvious mismatch between its divergent  parts 
and the corresponding counterterm.  
 
({\it v}) The fundamental STI employed in section II 
survives renormalization, simply because all 
counterterms necessary to render it finite are already furnished 
by the usual counterterms of the RFG Lagrangian.
This is, of course, a direct result of the basic assumption 
the the theory in the RFG is renormalizable:  
once all
counterterms have been supplied in the RFG, the STI
which is studied in the same gauge, will continue being valid.

({\it vi}) As has been explained in  \cite{Papavassiliou:2000az},
and as is obvious from the coincidence of the PT and BFG 
results,
the basic structure which appears nested inside 
the PT Green's functions is the 
high-order 
generalization of the vertex quantity 
$\Gamma_{\alpha\mu\nu}^{{\rm F}\,[0]}$. 
This quantity, 
to be denoted by
$\Gamma_{\alpha\mu\nu}^{{\rm F}\,[n]}(q,p_1,p_2)$ coincides 
with the all-order 
BFG Green's functions  with one background ($\tilde{A}$) 
and two quantum ($A$) gluons incoming, {\it i.e.} 
$\Gamma_{\tilde{A}_{\alpha} A_{\mu} A_{\nu}}^{[n]}(q,p_1,p_2) $.  
$\Gamma_{\alpha\mu\nu}^{{\rm F}\,[n]}(q,p_1,p_2)$
satisfies the following WI
\be
q^{\alpha} \Gamma_{\alpha\mu\nu}^{{\rm F}\,[n]}(q,p_1,p_2)
= \Pi_{\mu\nu}^{[n]}(p_1) - \Pi_{\mu\nu}^{[n]}(p_2) , 
\label{WI2B2}
\ee
which is the exact one-loop analog of 
the tree-level Ward identity
of Eq (\ref{WI2B}); indeed the RHS is the
difference of two conventional $n$-loop self-energies 
computed in the RFG.
Notice also that  Eq.\ (\ref{WI2B2}) dictates
that the ultraviolet-divergent part of  
$\Gamma_{\alpha\mu\nu}^{{\rm F}\,[1]}$ is proportional to 
$\Gamma_{\alpha\mu\nu}^{[0]}$ rather than
$\Gamma_{\alpha\mu\nu}^{{\rm F}\,[0]}$; had it been the other way around 
there would be no longitudinal ultraviolet-divergent 
pieces on the RHS of Eq.\ (\ref{WI2B2}). 
As has been explained in \cite{Papavassiliou:2000az}, this ``mismatch''
will generate the pieces which, in the background 
field method language, give rise to the
gauge-fixing renormalization of the vertices [see point ({\it vii}), below].
Clearly, due to 
the WI of  Eq.\ (\ref{WI2B2}), we must have 
${Z}_{3F} = {Z}_A$, where 
$Z_{3F}$ is the vertex renormalization constant for 
the $\Gamma_{\alpha\mu\nu}^{{\rm F}\,[n]}$.  

({\it vii}) After the rearrangements of the original counterterms 
(in the RFG),
 in such a way as to render the PT Green's functions finite, one  
should be able to verify 
that the resulting counterterms  are  in fact identical to those
obtained  when  carrying out   the background field method  
renormalization  program as explained by Abbott 
in the eighth paper of \cite{Dewitt:ub}, 
{\it i.e.} by  renormalizing  only  the  background
gluons,  the external  quarks,    the   coupling constant $g$, and the
quantum gauge-fixing parameter $\xi_Q$.
Thus, the relevant renormalization constants are given by
\be 
g_0 = Z_g g\,, \,\,\,\,\,\,\,\,
\tilde{A_0} = Z_{\tilde{A}}^{1/2} \tilde{A}\,, \,\,\,\,\,\,\,\,
\xi_Q^{0} = Z_{\xi_Q} \xi_Q \,,\,\,\,\,\,\,\,\, Z_{\xi_Q}=Z_A\,.
\ee 
The renormalization of $\xi_Q$ is necessary 
due to the fact that 
the longitudinal part of the quantum gluon propagator is not 
renormalized. As pointed out 
by Abbott, in the context of the background field method 
this step may be avoided if the calculation is carried 
out with an arbitrary $\xi_Q$ rather than the BFG $\xi_Q=1$. 
Of course, as we have seen,
the PT brings us effectively at $\xi_Q=1$; 
thus, when attempting to interpret the resulting counterterm 
from the background field method point of view, one should 
keep in mind that   
gauge-fixing parameter renormalization is necessary. 
The renormalization of $\xi_Q$ 
not only affects the propagator-lines, but also the 
longitudinal parts of the external vertices; it renormalizes
precisely the $\Gamma_P$ part, as can be seen from Eq.(\ref{secondec}).

All the above ingredients must be combined appropriately in order to 
demonstrate the renormalizability of the PT effective Green's functions; 
for the purposes of this paper we shall not pursue this 
point any further. 

\section{Discussion and Conclusions}

In this  article we have presented  in detail the  construction to all
orders  in perturbation theory  of three  basic PT  Green's functions,
namely  the off-shell  gluon self-energy,  the quark--anti-quark-gluon
vertex, and  the four-quark box.   
The PT procedure,
through its  very definition, is based on  the systematic exploitation
of  a  fundamental cancellation  between  the  self-energy and  vertex
diagrams  appearing  in the  amplitude  of  a  physical process.  This
cancellation  allows  for the  construction  of gauge-independent  and
gauge-invariant  effective  Green's  functions,  with the  variety  of
phenomenological  uses  outlined  in  the Introduction.   The  central
result of the  present paper is that this  crucial cancellation can be
carried  out  systematically  and   expeditiously  to  all  orders  by
appealing to the STI satisfied by a special four-point function, which
constitutes  a common kernel  to the  self-energy and  vertex diagrams
involved  in  the pinching  procedure.  Therefore,  all the  important
properties  of the  PT  Green's  functions, known  from  the one-  and
two-loop analysis, are valid to all orders.

As was first shown in \cite{Binosi:2002ft}, and in the present one in 
much more detail, 
the  known correspondence  between the  PT Green's
functions and  those calculated in the  BFG persists to  all orders. 
This fact  which provides  a very  convenient book-keeping  scheme  
for the actual calculation of  the former, in principle to  any desired order.
We emphasize that this  correspondence has been established through an
a-posteriori  comparison of  the PT  results, derived  in the  RFG, to
those of the BFG; all diagrammatic rearrangements leading to the latter
scheme,  and in particular  to its  very characteristic  ghost sector,
have proceeded dynamically, due to the appropriate exploitation of the
corresponding STIs.  It  would be clearly very interesting  to reach a
deeper understanding of  what singles out the value  $\xi_Q = 1$.  One
possibility would be to look  for special properties of the BFM action
at  $\xi_Q  =   1$  \cite{RDP};  an  interesting  3-d   example  of  a
field-theory, which,  when formulated  in the background  Landau gauge
($\xi_Q  =  0$),  displays  an  additional  (non-BRST  related)  rigid
super-symmetry, is given in \cite{Birmingham:1988ap}.

Despite the progress reported in the present article, 
various technical questions merit further study.
To begin with, the general construction of 
higher PT $n$-point functions with all legs off-shell 
(for example, the all-order
three-gluon vertex ($n=3$), whose one-loop derivation was presented in 
the first paper of \cite{Cornwall:1989gv}) is lacking for the moment.
In addition, our analysis has been restricted to the 
case of the linear covariant gauges, but it would be 
interesting to study what happens in the 
context of entirely different 
gauges, as for example is the case of
the non-covariant  axial gauges \cite{Dokshitzer:1980hw}.
These gauges present the additional complication that 
the  convenient Feynman gauge cannot be  reached a priori  by simply fixing
appropriately the value of the  gauge fixing parameter.
Our experience from  explicit one- and two-loops calculations 
(see for example 
\cite{Binosi:2001hy}, and the 
third paper of \cite{Nadkarni:1988ti}) is that 
the application of the usual PT algorithm  
leads to a vast number of cancellations, which dynamically projects one  
to the $g_{\mu\nu}$ part of the gluon propagator.
Thus, even if
one  uses a bare  gluon propagator  of the  general axial gauge form,
after the  aforementioned cancellations  have taken place  one arrives
effectively to the answer written in the RFG; it is an open question
whether this fact persists to all orders.
Needless to say, the generalization of the 
formalism developed here to the Electroweak sector 
of the Standard Model 
presents, as in the two-loop 
case \cite{Binosi:2002bs},
a significant technical challenge.
Furthermore, at the conceptual level it is unknown whether 
a formal definition of the PT Green's functions 
in terms of fundamental fields, encoding 
``ab initio'' their special properties, is possible.  
Finally, it would be interesting to explore possible connections 
with various related formalisms
\cite{Nielsen:1975ph,Bern:an,Catani:fe,DiVecchia:1996uq,Arnone:2002qi}.

\begin{acknowledgments} 
J.P. thanks R. Pisarski for his continuing encouragement, and 
F. del Aguila and N. Mavromatos for valuable comments. 
The work of D.B. is supported by the Ministerio of Educaci\'on, Cultura y
Deporte, Spain, under Grant DGICYT-PB97-1227, and 
the research of J.P. is supported by CICYT, Spain, under Grant AEN-99/0692.
\end{acknowledgments}


\begin{thebibliography}{99}
\bibitem{dis}
This discussion serves as a brief reminder; for further details 
the reader is 
referred to the rather extensive literature, 
cited throughout the present paper.
\bibitem{WI}
This WI is relevant of course when the external particles are quarks. 
An analogous WI is triggered if the  external particles 
are gluons or scalars.
\bibitem{com1} 
One may also think of these vertices as interaction vertices   
of composite operators; however, the nature 
and possible role (if any)
of such operators in the present context, is unclear to us.
\bibitem{unitary}
A classic counter-example is the unitary gauge of the 
Standard Model, 
where the entire $S$-matrix is renormalizable, whereas the individual 
Green's functions computed in this gauge are non-renormalizable.
\end{thebibliography}
\end{document}